%% file: paper.tex
\documentclass[conference]{IEEEtran}
\usepackage{amsmath,amssymb,graphicx,hyperref,times,balance}
\usepackage[tight]{subfigure}
\IEEEoverridecommandlockouts

\begin{document}
\title{Greedy Forwarding in Dynamic Scale-Free Networks Embedded in Hyperbolic Metric Spaces}
\author{Fragkiskos Papadopoulos, Dmitri Krioukov,  Mari{\'a}n Bogu{\~n}{\'a},  and Amin Vahdat
\thanks{F.~P.~is with the Department of Electrical and Computer
Engineering, University of Cyprus. D.~K.~is with the Cooperative Association for Internet Data Analysis (CAIDA),
and A.~V.\ is with the Department of Computer Science and Engineering,
University of California, San Diego (UCSD). M.~B.~is with Department de F{\'\i}sica Fonamental, Universitat de
Barcelona. The work was supported by NSF CNS-0722070 and Cisco Systems,
and done while the first author was with CAIDA.}}
\maketitle
\begin{abstract}
  We show that complex (scale-free) network topologies
  naturally emerge from hyperbolic metric spaces.
  Hyperbolic geometry facilitates maximally efficient
  greedy forwarding in these networks. Greedy forwarding
  is topology-oblivious. Nevertheless, greedy packets find their
  destinations with $100\%$ probability following almost optimal
  shortest paths. This remarkable efficiency sustains
  even in highly dynamic networks. Our findings suggest that forwarding information through complex networks,
   such as the Internet, is possible without the overhead of existing routing protocols, and may also
   find practical applications in overlay networks for tasks such as application-level
   routing, information sharing, and data distribution.
\end{abstract}
\input{introduction}
\input{related}
\input{static}
\input{dynamic}
\input{discussion}
\input{conclusion}
\input paper.bbl

\balance
\input{appendix}
\end{document}

%% file: introduction.tex
\section{Introduction}
\label{sec:introduction}

Routing information is the most basic and, perhaps, the most
complicated function that networks perform. Conventional wisdom states
that to find paths to destinations through the complex network maze,
nodes must collectively discover a current
state of the network topology by exchanging information about the status of
their connections to other nodes. This communication overhead is
considered one of the most serious scaling limitations of our primary
communication technologies today, including the
Internet \cite{iab-raws-report-phys} and emerging wireless and sensor
networks \cite{NaGro05}. Finding intended communication targets in
other networks, such as P2P overlays, relies
on flooding-based mechanisms, random
walks, and other techniques,
whose efficiency may be unpredictable,
and overhead costs unbounded \cite{p2psurvey}.

However, many networks in nature can somehow ``route traffic''
efficiently. That is, nodes in these networks can efficiently find intended
communication targets even though they do not possess any global view
of the system. Milgram's 1969 experiment \cite{TraMi69} showed a
classic demonstration of this effect.  Milgram asked some random
individuals---sources---to send a letter to a specific person---the
destination, described by name, occupation, age, and city of
residence. The sources were asked to pass the letter to friends chosen
to maximize the probability of the letter reaching its
destination. The results were surprising: many of the letters reached
their destination by making only a small number of hops, even though
nodes had no global knowledge of the human acquaintance network
topology, except their local connections and some characteristics
(e.g., occupation, age, city of dwelling) of their connections.

Much later, Jon Kleinberg offered the first popular explanation of
this surprising effect \cite{kleinberg00-nature}. In his model, each
node, in addition to being part of the graph representing the global
network topology, resides in a coordinate space---a grid embedded in
the Euclidean plane. The coordinates of a node in the plane, its
address, abstracts information about the destination in Milgram's
experiments. Each node knows: 1)~its coordinates; 2)~the coordinates
of its neighbors; and 3)~the coordinates of the destination written on
the packet. Given these three pieces of information, the node can
route greedily by selecting its direct neighbor closest to the
destination in the plane.

Clearly, the described greedy forwarding strategy can be efficient
\emph{only if} the network topology is in some way congruent with the
underlying space.  But the Kleinberg model does not (try to) reproduce
the basic topological properties of social networks through which
messages were traveling in Milgram's experiments.  For instance, the
model produces only $k$-regular graphs while social networks, the
Internet, and many other complex networks~\cite{newman03c-review} are
known to be {\it scale-free}, meaning that: i)~the distribution $P(k)$
of node degrees $k$ in a network follows power laws $P(k)\sim
k^{-\gamma}$ with exponent $\gamma$ often lying between $2$ and $3$;
and ii)~the network has strong clustering, i.e., a large number of
triangular subgraphs \cite{DorMen-book03}.

Our work follows Kleinberg's formalism. We assume that nodes in
complex networks exist in some spaces that underlie the observed
network topologies. We call these spaces {\it hidden metric
spaces}. The observed network topology is coupled to the hidden
space geometry in the following way: a link between two nodes in the
topology exists with a certain probability that depends on the
distance between the two nodes in the hidden geometry. A
plausible explanation for the Kleinberg model's inability to
naturally produce scale-free topologies is that the spaces hidden
beneath such topologies are not Euclidean planes.

The primary contribution of this paper is the demonstration that a
simple mechanism of network growth in a \emph{hyperbolic hidden
metric space} naturally leads to the emergence of scale-free
topologies.  One attractive property of such topologies is that greedy
forwarding using node coordinates in the hyperbolic space
results in 100\% reachability with nearly optimal path lengths even
under dynamic network conditions, with link failures and node arrivals
and departures. Most importantly, nodes do not change their coordinates
upon network topology changes. Therefore, nodes do not have to exchange
any routing information even in dynamic networks. Our work thus paves a path to
nearly optimal forwarding in complex networks, such as the Internet,
without expensive and brittle routing protocols.
Our results may also
find practical applications in overlay networks for tasks such
as application-level routing, information sharing, and data
distribution.

The rest of the paper is organized as follows. In Section~\ref{sec:related}
we discuss related work. In Section~\ref{sec:static}
we reveal the connection between scale-free network topologies and
hyperbolic geometries, and present a simple model that builds
scale-free networks using such geometries. This model builds a network
as a whole at once. In the same section we
demonstrate the remarkable efficiency and robustness
of greedy forwarding in dynamic scenarios with link failures. Since in
many practical applications nodes may arrive to the system gradually,
in Section~\ref{sec:dynamic} we extend our model for scale-free networks
that grow in hyperbolic spaces. We
demonstrate that greedy forwarding strategies are still extremely
efficient, even under highly dynamic network conditions with nodes
randomly arriving and departing the system. We discuss practical
applications of our findings in Section~\ref{sec:applications}, and
conclude with directions for future research in Section~\ref{sec:conclusion}.

%% file: related.tex
\section{Related work}
\label{sec:related}

The most relevant earlier work is the groundbreaking result by Robert
Kleinberg, who shows in~\cite{kleinberg07infocom} how any given graph
can be embedded in the hyperbolic plane such that greedy forwarding
can achieve $100\%$ reachability. However, to construct the embedding
one needs to know the graph topology in advance. R. Kleinberg's work
has been recently complemented by the work in
\cite{crovella09infocom}, where the authors propose a simple technique
for online embedding of any given graph in the hyperbolic plane. The
graph can also be dynamic, in the sense that once the initial graph is
embedded, its topology can change, and greedy forwarding can still
maintain $100\%$ reachability.

While \cite{kleinberg07infocom} and \cite{crovella09infocom} show how
any \emph{given} graph can be embedded in a hyperbolic space so that
greedy forwarding maintain $100\%$ reachability, here we approach the problem from the
opposite direction. We fix the hyperbolic space and construct graphs in it
in the simplest possible manner. We show that the resulting graphs are
not \emph{any} graphs but \emph{scale-free} graphs, \emph{naturally}
congruent with underlying hyperbolic geometry. That is, we do nothing
to enforce these graphs to be scale-free; their scale-free topology
emerges naturally as a consequence of underlying hyperbolic
geometry. Because of this congruency, greedy forwarding
strategies are maximally efficient, even in presence of network dynamics.
In terms of practical applications, the studies in \cite{kleinberg07infocom} and
\cite{crovella09infocom} are more suitable for cases where the
initial network topology is already globally known, and where
the costs associated with such global knowledge are low,
while our work here is
more applicable to cases where networks are formed dynamically, where
we do not know their exact topology in advance, such as in overlay
network applications, \cite{p2psurvey, overlay_routing}, and to cases
where such global topology awareness may be prohibitive in terms of
associated routing overhead costs. See Section \ref{sec:applications}
for a discussion of potential applications.

As mentioned earlier, the first popularization of greedy routing as a
mechanism that might be responsible for efficient forwarding ``in the
dark'', i.e., without the knowledge of network topology, is due to Jon
Kleinberg~\cite{kleinberg00-nature}. A vast amount of literature
followed this seminal work, as reviewed in~\cite{kleinberg06-review}.
Other works, dealing with hyperbolic geometry in the network context,
include~\cite{ShaTa04b,KraLe06,AbBa07,JoLoBo08}. No earlier work has considered
\emph{hidden} hyperbolic geometries,
which can be used to efficiently guide
the forwarding process on complex networks.

Finally, there has been a great deal of research trying to explain the
scale-free structure of complex networks \cite{BoSchu02-book}, among
which preferential attachment \cite{BarAlb99} appears to be the most
popular. However, no existing effort has considered hidden hyperbolic
geometry as a possible explanation.

%% file: static.tex
\section{Scale-free networks and hyperbolic spaces}
\label{sec:static}

In this section we first provide high-level intuition behind the
connection between scale-free network topologies and hyperbolic
geometries. We then proceed by presenting a simple model where
scale-free topologies naturally emerge from such geometries, and
demonstrate the remarkable efficiency of greedy forwarding strategies
that use these geometries.

\subsection{Intuition}

The main metric property of hyperbolic geometry that we use in this
paper is the exponential expansion of space. For example, in the
hyperbolic plane, which is the two-dimensional hyperbolic space of
negative curvature $-1$, the length of a circle and the area of a disc
of radius $R$ are $2\pi\sinh R$ and $2\pi(\cosh R -1)$, both growing
as $\sim e^R$ with $R$.~\footnote{In this paper, symbols `$\sim$' and
  `$\approx$' mean, respectively, {\it proportional to} and {\it
    approximately equal}.}  The hyperbolic plane is thus metrically
equivalent to an $e$-ary tree, i.e., a tree with the average branching
factor equal to $e$. Indeed, in a $b$-ary tree, the analogies of the
circle length or disc area are the number of nodes at distance exactly
$R$ or not more than $R$ hops from the root. These numbers are
$(b+1)b^{R-1}$ and $((b+1)b^R-2)/(b-1)$, both growing as $\sim
b^R$. Informally, hyperbolic spaces can therefore be thought of as
``continuous versions'' of trees.  Other properties of hyperbolic
geometry can be found in various (text)books, e.g.,
\cite{BridsonHaefliger99-book}.

To see why this exponential expansion of hidden space is intrinsic to
scale-free networks, observe that their topology represents the
structure of connections or interactions among distinguishable,
heterogeneous elements abstracted as nodes.  The heterogeneity implies
that nodes can be somehow classified, however broadly, into a
taxonomy, i.e., nodes can be split into large groups consisting of
smaller subgroups, which in turn consist of even smaller subsubgroups,
and so on. The relationships between such groups and subgroups, called
communities~\cite{GiNe02}, can be approximated by tree-like
structures, in which the distance between two nodes estimates how
similar they are~\cite{WatDoNew02,ClMo08}. The smaller the distance,
the more similar the two nodes are, and the more likely they are
connected. Importantly, the node classification hierarchy need not be
strictly a tree.  Approximate ``tree-ness,'' which can be formally
expressed solely in terms of the metric structure of a
space~\cite{Gromov07-book}, makes the hidden space hyperbolic.~\footnote{We
  call the space \emph{hidden} to emphasize that the distance between
  two nodes in it is a measure of how similar they are; it is
  \emph{not} their shortest path distance in the observable network
  graph as in \cite{ShaTa04b,KraLe06,AbBa07,JoLoBo08}.}

\subsection{Models of scale-free networks in hyperbolic spaces}

We now put our intuitive considerations to qualitative grounds. We
want to see what network topologies emerge in the simplest possible
settings involving hidden hyperbolic metric spaces. Specifically, we
use the following strategy to formulate a network model.  We specify:
1)~the hyperbolic space; 2)~the distribution of nodes in it, i.e., the
node density; and 3)~the connection probability as a function of the
hyperbolic distance between nodes, i.e., we connect a pair of nodes
located at hyperbolic distance $d$ with some probability $p(d)$.

The simplest hyperbolic space is the two-dimensional hyperbolic plane
of constant negative curvature $-1$ we discussed earlier.  The
simplest way to place $N$ nodes on the hyperbolic plane is to
distribute them uniformly over a disc of radius $R$. The
hyperbolically uniform node density implies that we assign the angular
coordinates $\theta\in[0,2\pi]$ to nodes with the uniform density
$f(\theta)=1/(2\pi)$, while the density for the radial coordinate
$r\in[0,R]$ is exponential $f(r)=\sinh r/(\cosh R - 1) \approx e^{r-R}
\sim e^r$, as the circle length at distance $r$ from the disc center
is $2\pi\sinh r$ (vs.\ $f(r) \sim r$ in the Euclidean plane, where the
circle length is $2\pi r$). We can also generalize the model by
distributing nodes non-uniformly on the disc using:
\begin{equation}
\label{eq:rho(r)}
f(r)= \frac{\alpha \sinh \alpha r}{\cosh \alpha R - 1} \approx \alpha e^{\alpha(r-R)} \sim e^{\alpha r},
\end{equation}
with $\alpha=1$ corresponding to the hyperbolically uniform node density.

The simplest connection probability we could think of is the step
function $p(d)=\Theta(R-d)$, meaning that we connect a pair of nodes
with polar coordinates $(r,\theta)$ and $(r',\theta')$ by a link only
if the hyperbolic distance between them is $d \leq R$, where $d$ is
given by the hyperbolic law of cosines: $\cosh
d=\cosh{r}\cosh{r'}-\sinh{r}\sinh{r'}\cos{\Delta\theta}$, with
$\Delta\theta=|\theta-\theta'| \mod \pi$.
The following theorem states that the node degree distribution in the
resulting network is a power law.  \newtheorem{thm}{\textbf{Theorem}}
\begin{thm}
\label{thm:one}
The described model produces graphs with the power law node degree
distribution:
\begin{equation}
\label{eq:p(k)}
P(k) \sim k^{-\gamma}, \quad \text{with}\; \gamma =
\begin{cases}
2\alpha+1&\text{if $\alpha \geq \frac{1}{2}$},\\
2&\text{if $\alpha \leq \frac{1}{2}$}.
\end{cases}
\end{equation}
\begin{proof}
  We first compute the average degree $\bar{k}(r)$ of nodes located at
  distance $r$ from the disc center. Such nodes are connected to all
  nodes in the intersection area of the two discs of the same radius
  $R$, one in which all nodes reside, and the other centered at
  distance r from the center of the first disc. Its approximate,
  simplified expression is:
\begin{equation}
\label{eq:k(r)}
\bar{k}(r) \approx N\left\{\frac{2}{\pi}\frac{\alpha}{\alpha-\frac{1}{2}} e^{-\frac{1}{2}r} +
\left(1-\frac{2}{\pi}\frac{\alpha}{\alpha-\frac{1}{2}}\right)e^{-\alpha r}\right\},
\end{equation}
where the limit $\alpha\to \frac{1}{2}$ is $\bar{k}(r) \rightarrow N
\left(1+\frac{r}{\pi}\right) e^{-\frac{1}{2}r}$.~\footnote{We
    omit the intermediate calculations for brevity. All the
    omitted calculations can be found in the technical report
  \cite{tech_report}.}
 From Equation (\ref{eq:k(r)}) we see
that $\bar{k}(r)$ decreases exponentially, i.e., $\bar{k}(r) \sim
e^{-\beta r}$, with $\beta=\frac{1}{2}$ if $\alpha \geq \frac{1}{2}$
and $\beta=\alpha$ if $\alpha \leq \frac{1}{2}$. Therefore, $\bar{r}(k) \sim
-\frac{1}{\beta} \ln k$. Given $f(r)$ from Equation (\ref{eq:rho(r)}),
it is easy to see that $P(k) \approx f(\bar{r}(k))\,|\bar{r}'(k)|\sim k^{-\gamma}$
with $\gamma=\frac{\alpha}{\beta}+1$. Thus, $\gamma=2\alpha+1$ if
$\alpha \geq \frac{1}{2}$ and $\gamma=2$ if $\alpha \leq \frac{1}{2}$.
\end{proof}
\end{thm}

We thus see that by changing $\alpha$, which according to our tree
analogy regulates the average branching factor of the hidden tree-like
hierarchy, we can construct power-law graphs with any exponent $\gamma
\geq 2$, as observed in a majority of known complex networks,
including the Internet \cite{DorMen-book03}.

The average node degree $\bar{k}=\int_0^R  \bar{k}(r) f(r) dr$, is:
\footnotesize
\begin{eqnarray}
\nonumber \bar{k} \approx N \frac{\left\{2\alpha^2 e^{-\frac{1}{2}R}+\left[\left((\pi-2)\alpha^3-(\pi-1)\alpha^2+\frac{\alpha\pi}{4}\right)R-2\alpha^2\right]e^{-\alpha R}\right\}}
{\pi\left(\alpha-\frac{1}{2}\right)^2},
\label{eq:kbar}
\end{eqnarray}
\normalsize where the limit $\alpha \rightarrow \frac{1}{2}$ is
$\bar{k} \rightarrow N
\frac{R}{2}\left(1+\frac{R}{2\pi}\right)e^{-\frac{1}{2}R}$. Therefore,
given a target $\bar{k}$, a target exponent $\gamma$, which is related
to $\alpha$ via Equation (\ref{eq:p(k)}), and the number of nodes $N$,
the right value for the hyperbolic disc radius $R$ is (numerically)
computed by the above formula.  From the formula, we can also see that
the relationship between $R$ and $N$ is approximately logarithmic,
i.e., $R \sim \ln N$.

Theorem \ref{thm:one} states that the degree
distribution is a power law, but it does not give its exact
expression. Skipping calculations, this expression is:
\begin{equation}
\label{eq:p(k)_exact}
P(k)=2\alpha\xi^{2\alpha}\frac{\Gamma\left(k-2\alpha,\xi\right)}{k!},
\end{equation}
where $\xi=\bar{k}(2\alpha-1)/(2\alpha)$ and $\Gamma$ is the incomplete gamma function.

Our networks also possess strong clustering. Strong clustering, or
large numbers of triangles in generated networks, is a simple
consequence of the triangle inequality in the metric space. Indeed, if
node $a$ is close to node $b$ in the plane, and $b$ is close to a
third node $c$, then $a$ is also close to $c$ because of the triangle
inequality. Since all three nodes are close to each other, links
between all of them forming triangle $abc$ exist.

\begin{figure*}
    \centerline{
        \subfigure[Degree distribution $P(k)$.]{\includegraphics[width=1.8in]{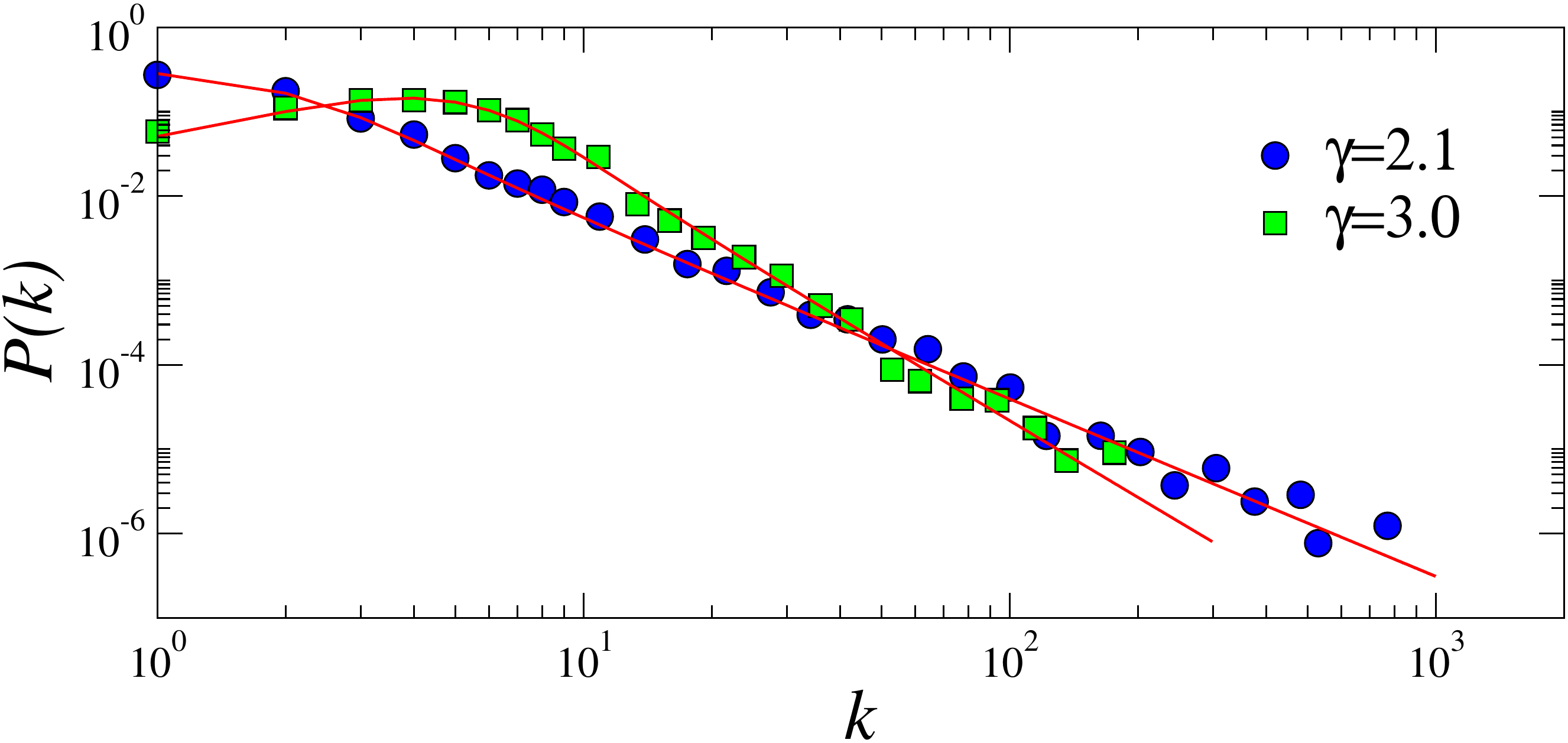}}\hfill
        \subfigure[Average clustering $\bar{c}(k)$.]{\includegraphics[width=1.8in]{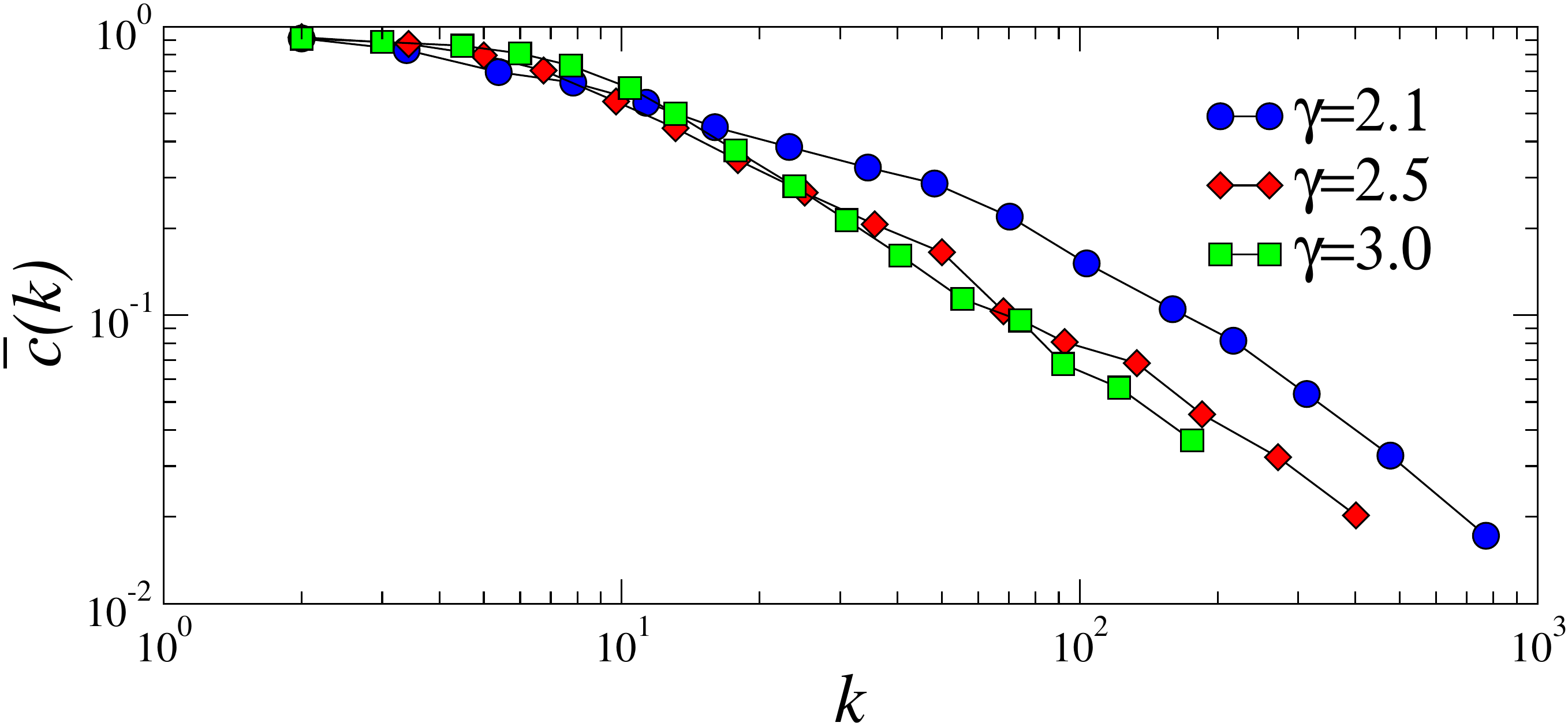}}\hfill
        \subfigure[$P(k)$, Model vs. Internet.]{\includegraphics[width=1.8in]{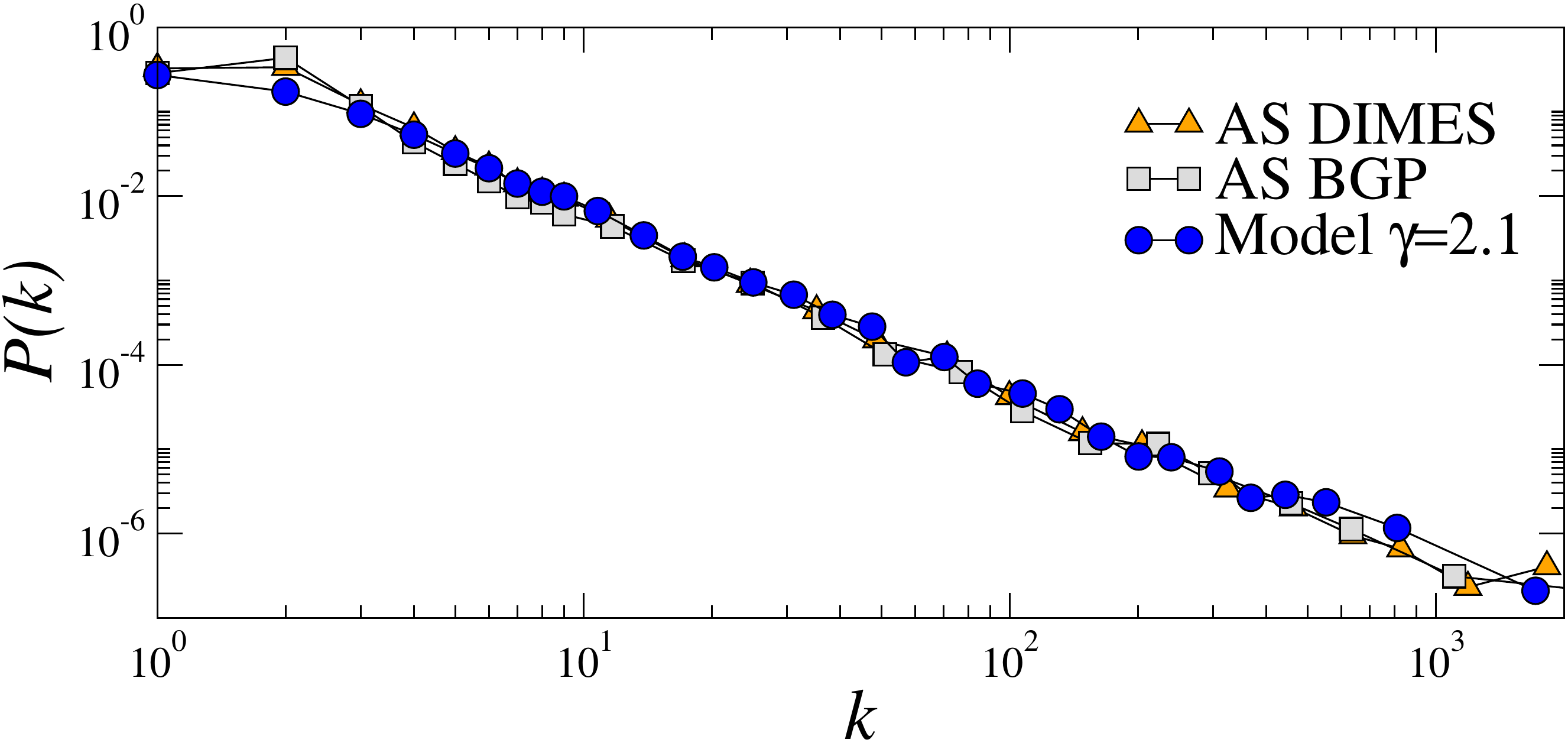}}\hfill
        \subfigure[$\bar{c}(k)$, Model vs. Internet.]{\includegraphics[width=1.8in]{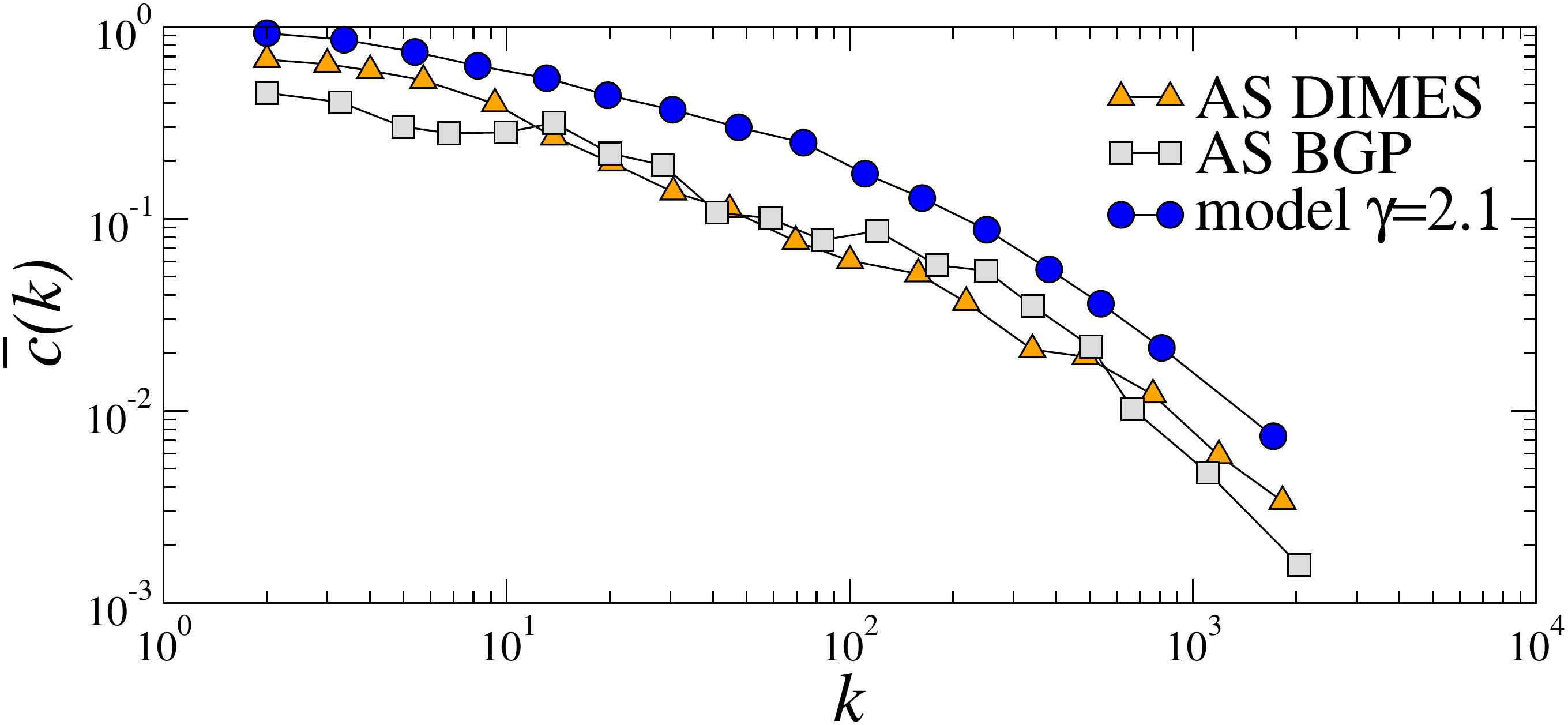}}
    }
    \caption{The first two plots show the degree distribution $P(k)$,
    and average clustering $\bar{c}(k)$ of $k$-degree nodes. The degree distribution for $\gamma=2.5$ is
    not shown for clarity. Solid lines in the first plot  are the theoretical
    prediction given by Equation (\ref{eq:p(k)_exact}).
    The last two plots show the same
    statistics for simulated networks with $\gamma=2.1$ vs.\ AS topologies from
    RouteViews BGP tables~\protect\cite{routeviews} and DIMES traceroute
    data~\protect\cite{dimes}.}
     \label{fig:static_stats}
\end{figure*}

In Figures \ref{fig:static_stats}(a) and \ref{fig:static_stats}(b) we
show, in log-log scale, the degree distribution $P(k)$ and average
clustering as a function of the node degree $\bar{c}(k)$
\cite{WatStr98}, for modeled networks with $N=10000$ and
$\bar{k}=6.5$. We observe agreement between simulation results and the
analytical prediction for the degree distribution in Equation
(\ref{eq:p(k)_exact}). In Figures \ref{fig:static_stats}(c) and
\ref{fig:static_stats}(d) we compare the same statistics between the
modeled networks with $\gamma=2.1$, and the AS Internet
topologies from RouteViews BGP tables \cite{routeviews} and DIMES
traceroute data \cite{dimes}. The degree distribution in our networks
is remarkably close to the empirical AS degree distribution.
The shape of the clustering curve $\bar{c}(k)$
in our networks is similar to the Internet's. In \cite{curvtemp} we
also show how clustering can be matched exactly.

\subsection{Greedy forwarding}

We now evaluate the performance of greedy forwarding (GF) strategies
on our modeled networks. A node's address is its hyperbolic
coordinates, and each node knows only its own address, the addresses
of its neighbors, and the destination address written in the
packet. GF forwards a packet at each hop to the neighbor closest to
the destination in the hyperbolic space. We present simulation results
for two simple forms of GF, {\it original} (OGF) and {\it modified}
(MGF).  The OGF algorithm drops the packet if the current hop is a
{\em local minimum}, meaning that it does not have any neighbor closer
to the destination than itself. The MGF algorithm excludes the current
hop from any distance comparisons, and finds the neighbor closest to
the destination. The packet is dropped only if this neighbor is the
same as the packet's previous hop.  We report the following metrics:
(i)~the percentage of successful paths, $p_s$, which is the proportion
of paths that reach their destinations; and (ii)~the average and
maximum stretch of successful paths, denoted by $\bar{s}$ and
$\text{max}(s)$ respectively. The stretch is defined as the ratio
between the hop-lengths of greedy paths and the corresponding shortest
paths in the graph.

We initially focus on static networks, where the network topology does
not change, and then emulate network topology dynamics by randomly
removing one or more links from the topology. As before, we fix the
target number of nodes in the network to $N=10000$ and its average
degree to $\bar{k}=6.5$, which roughly matches the Internet's AS
topology.  For each generated network, we extract the Giant Connected
Component (GCC), and perform GF between $10000$ random
source-destination pairs.
\begin{figure*}
    \centerline{
        \subfigure[Success ratio. (Static networks)]{\includegraphics[width=1.8in]{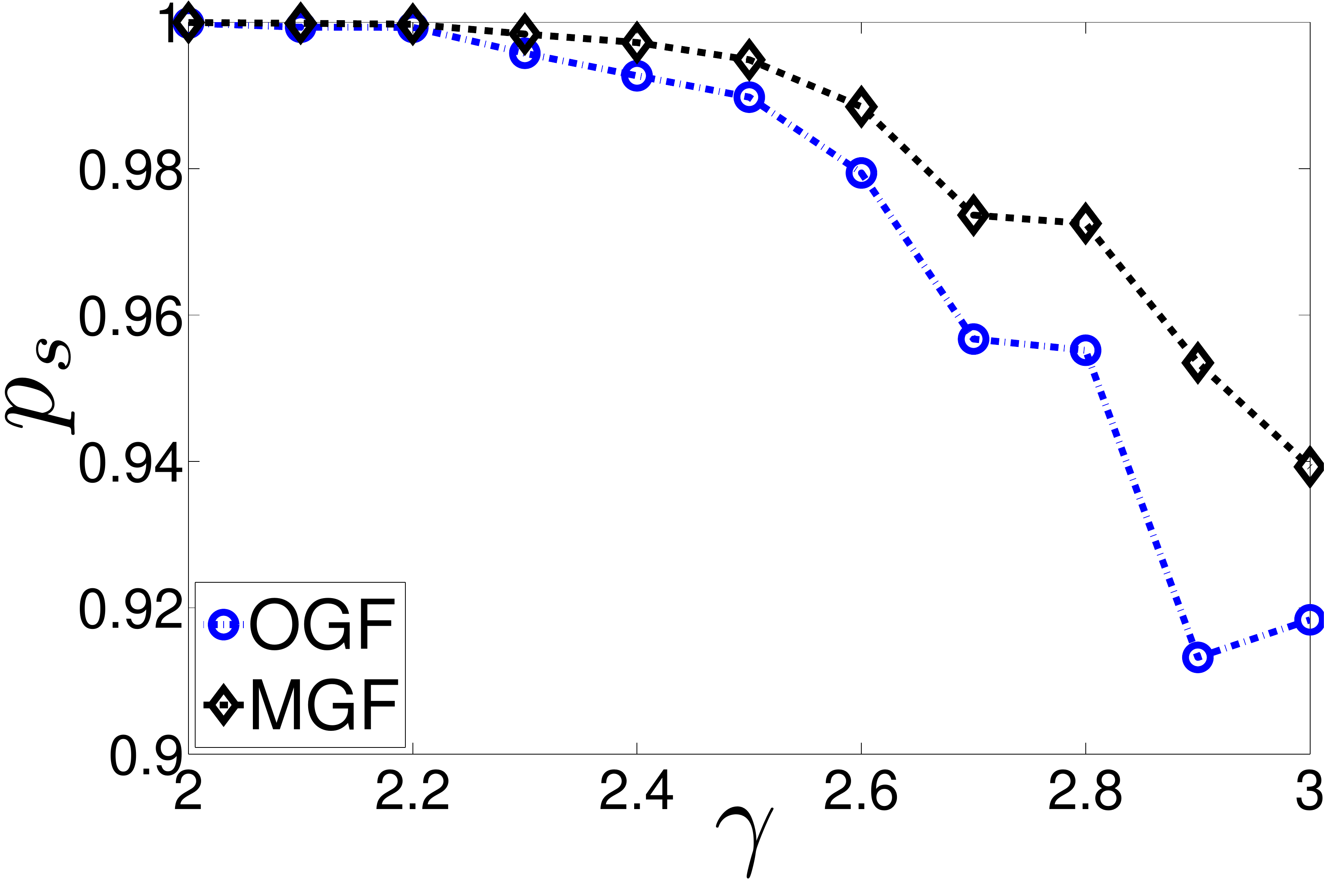}}\hfill
        \subfigure[Stretch. (Static networks)]{\includegraphics[width=1.8in]{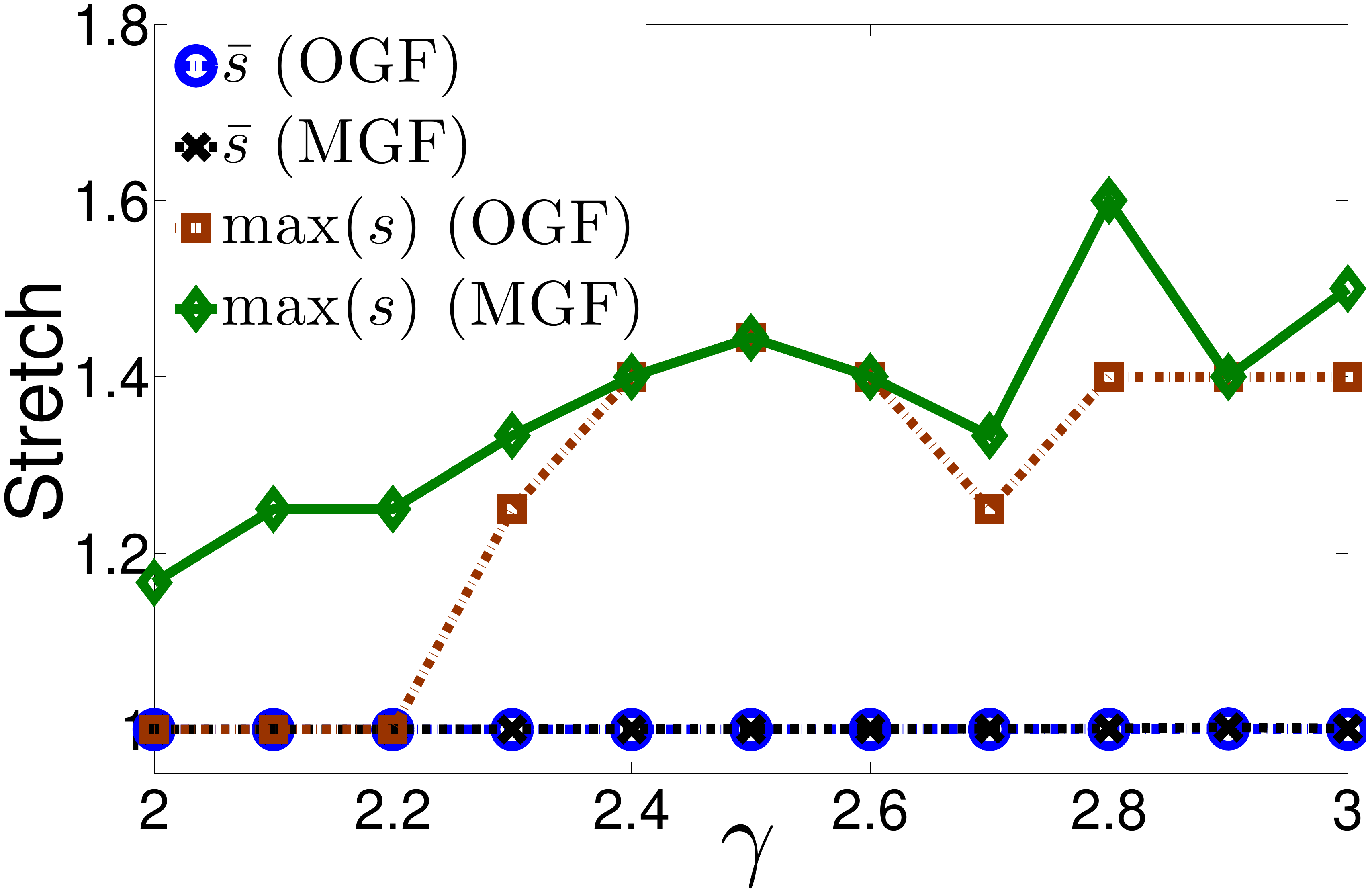}}\hfill
        \subfigure[Scenario 1. (Link failures)]{\includegraphics[width=1.8in]{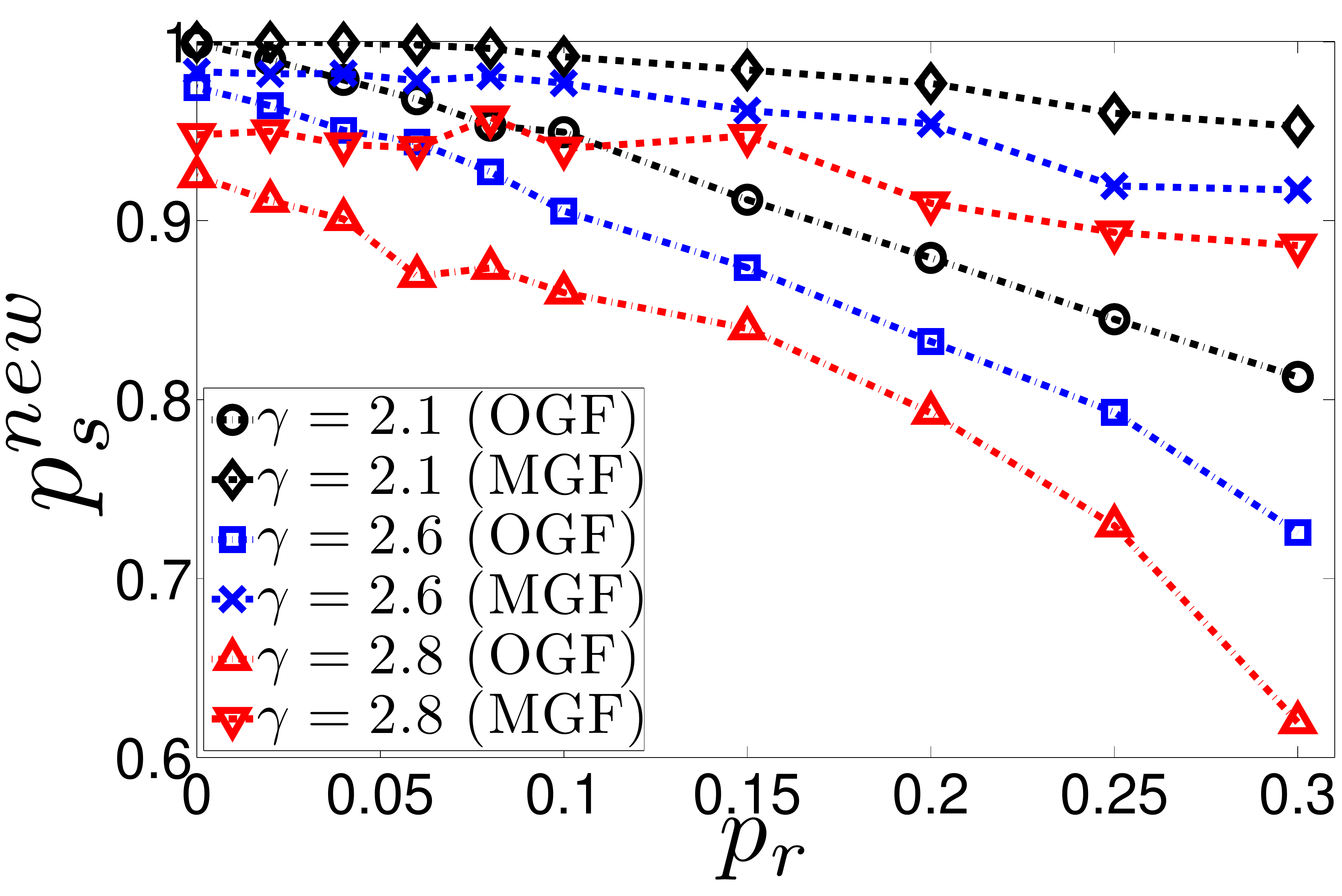}}\hfill
        \subfigure[Scenario 2. (Link failures)]{\includegraphics[width=1.8in]{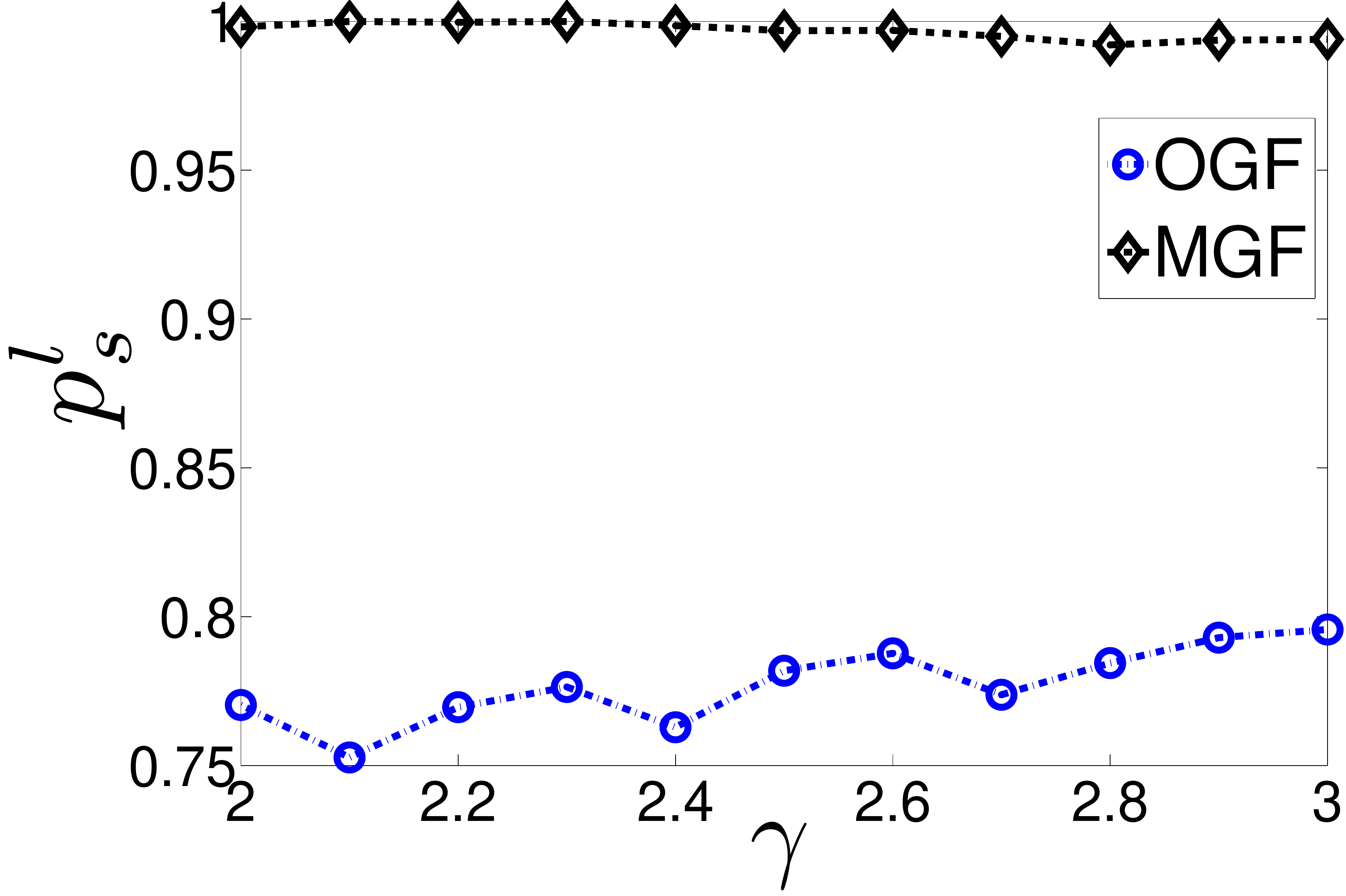}}
    }
    \caption{Performance of greedy forwarding (GF).
    \label{fig:greedy_performance}
    }
\end{figure*}

\begin{figure}[!ht]
  \begin{center}
    \includegraphics[width=2.4in]{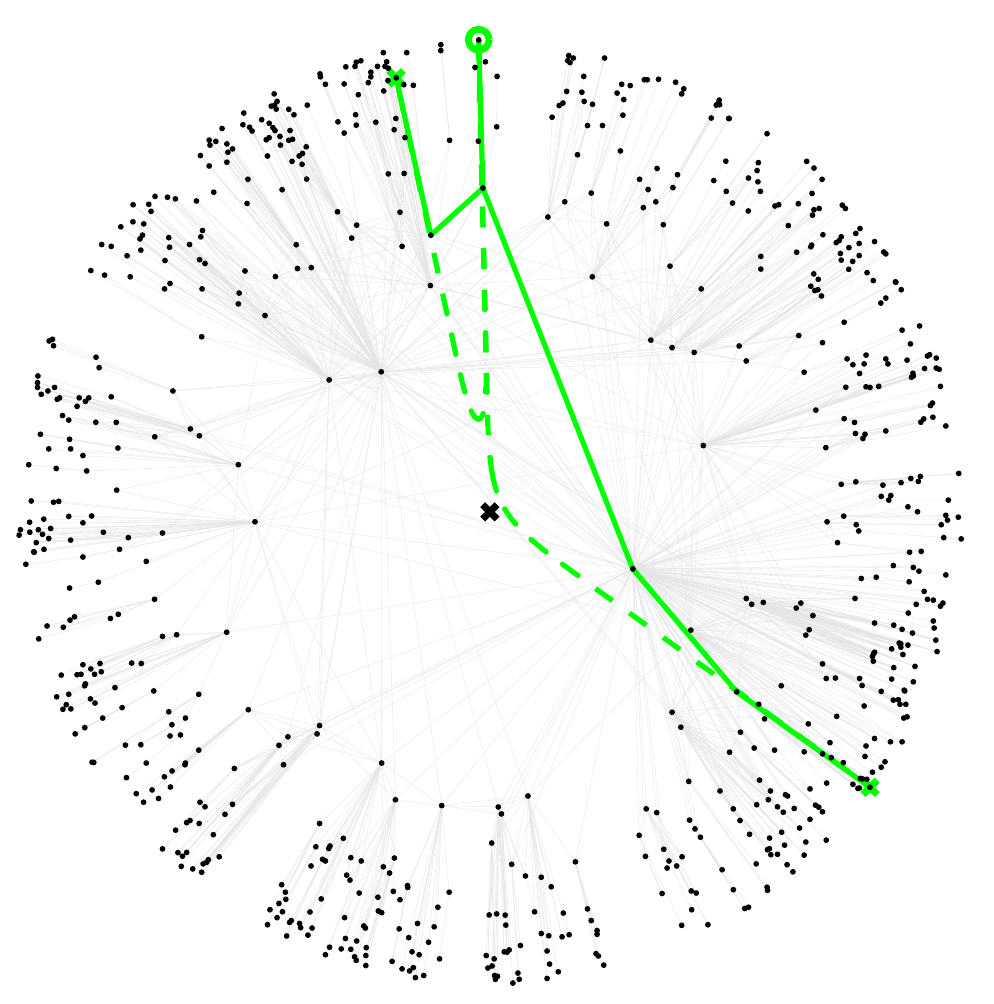}
    \caption{Visualization of a modeled network embedded in the hyperbolic plane, and greedy forwarding in it.
    The figure shows two hyperbolically straight lines, i.e., geodesics, the dashed curves, vs.\ the greedy paths, the solid lines, between
    the same source-destination nodes. The source is the top circled node, and the destinations are marked by crosses. The geodesics and greedy
    paths are approximately congruent as they follow the same pattern, first going to the disc center, and then veering off towards the destination.
    }
    \label{fig:graph-instance}
  \end{center}
\end{figure}

{\bf Static networks.} Figures \ref{fig:greedy_performance}(a) and
\ref{fig:greedy_performance}(b) show the results for static networks
of different degree exponent $\gamma$. We see that the success ratio
$p_s$ increases and the stretch decreases as we decrease $\gamma$ to
$2$. For example, for $\gamma=2.1$, i.e., equal to $\gamma$ observed
in the AS Internet, OGF and MGF yield $p_s=0.99920$ and $p_s=0.99986$,
with the OGF's maximum stretch of $1$, meaning that {\em all greedy
  paths are shortest paths}.  In summary, GF is exceptionally
efficient in static networks, especially for the small $\gamma$'s
observed in the vast majority of complex networks
\cite{DorMen-book03}. The two GF algorithms yield high success ratios
close to $1$ and optimal (or almost optimal) path lengths, i.e.,
stretch close to $1$.
The reason for this remarkable GF performance is the congruency
between the network topology and the underlying hyperbolic geometry,
as visually demonstrated in Figure \ref{fig:graph-instance}.

{\bf Link failures.} We next study the GF performance in dynamic
scenarios with link failures. We consider the following two
link-failure scenarios. In \underline{Scenario 1} we remove a
percentage $p_r$, ranging from $0\%$ to $30\%$, of all links in the
network, recompute the GCC, and compute the new success ratio
$p^{new}_{s}$. In \underline{Scenario 2}, we provide a finer-grain view
focusing on paths that used a removed link. We remove one link from
the network, recompute the GCC, and find the percentage $p^{l}_{s}$ of
successful paths, only among those previously successful paths that
traversed the removed link and belong to the new GCC. We repeat the
procedure for $1000$ random links, and report the average value for
$p^{l}_{s}$.  Figures \ref{fig:greedy_performance}(c) and
\ref{fig:greedy_performance}(d) present the results. We see that for
small $\gamma$'s, the success ratio $p^{new}_{s}$ remains remarkably
high, for all meaningful values of $p_{r}$.  For example, MGF on
networks with $\gamma=2.1$ and $p_{r} \leq 0.1$, yields $p^{new}_{s} >
0.99$.  Note that the simultaneous failure of $10\%$ of the links in
networks like the Internet is a rare catastrophe, but even in this
case GF is still efficient. The percentage $p^{l}_{s}$ of MGF paths
that used a removed link and that found a by-pass after its removal is
also remarkably high, close to $100\%$ for small $\gamma$'s. We
do not present the stretch results for brevity. In both
scenarios, and for all $\gamma$'s, the average stretch
remains remarkably low, below $1.1$.

In summary, GF is not only efficient in static networks, but its
efficiency is also remarkably robust with respect to network topology
dynamics. Thanks to high path diversity in scale-free networks,
there are many shortest paths,
disjoint over some links or nodes, between the same source
and destination, which all closely follow their geodesics.
Link removals affect some shortest paths, but others remain,
and greedy forwarding can use the underlying hyperbolic
``guidance system'' to find them.

{\bf Greedy forwarding modifications.}
Although the success ratios in scale-free networks with small
$\gamma$'s are extremely close to $1$, they are not exactly $1$.
However, since the performance of our simple GF strategies deteriorates
quite slowly with increasing network dynamics, we can expect that
simple GF modifications can
achieve $100\%$ success ratio and small
stretch even in more extreme dynamic network conditions than those
we have studied above. We will check our expectations in the next
section, in scenarios with random node arrivals and departures.

%% file: dynamic.tex
\section{Networks growing in hyperbolic spaces}
\label{sec:dynamic}

The model we presented in the previous section generates a whole
network at once.  However, in many applications, the network topology
is formed by nodes gradually arriving over time. In this section,
we extend our model for scale-free
networks that grow in hyperbolic spaces. We then demonstrate the
remarkable efficiency of greedy forwarding in highly dynamic
conditions, with nodes randomly arriving and departing the system.

\subsection{Growing model}

We assume that the network initially consists of $0$ nodes. We
number each arriving node by its order of arrival. We do
not consider node departures for now. A new node $i \geq 1$ that arrives to the
system needs to know: (i) the current number of nodes in the network,
including itself, $N(i)=i$; (ii) a system pre-specified parameter
$\alpha$ for the node radial density, which as before, will determine
the exponent of the degree distribution $\gamma$; and (iii) a system
pre-specified constant $c$, which will determine the average node
degree as will be explained below.  Then, to connect to the network,
the node performs the following operations inspired by the model in
Section \ref{sec:static}:
\begin{itemize}
\item[i.] select an angular coordinate $\theta$ uniformly distributed in $[0,2\pi]$;
\item[ii.] compute the current hyperbolic disk radius $R(i)$ according to $R(i)=\frac{1}{\alpha} \ln\frac{i}{c}$, i.e., $i=ce^{\alpha R(i)}$;
\item[iii.] select a radial  coordinate $r \in [0, R(i)]$, according
  to the probability density  function $f(r|R(i))=\frac{\alpha
    \text{sinh}(\alpha r)}{\text{cosh}(\alpha R(i))-1}\approx \alpha
  e^{\alpha(r-R(i))}$;
\item[iv.] connect to every node $1 \leq j < i$, already in the
  network, for which the hyperbolic distance to it, denoted by
  $d_{ij}$,  satisfies $d_{ij} \leq R(i)$.~\footnote{To
    maintain graph connectedness at any time instance, node $i$ having
    such coordinates that $d_{ij} > R(i)~\forall j$ can randomly
    re-select new coordinates until $d_{ij} \leq R(i)$ for at least one
    $j$.}
\end{itemize}

To build the network in a fully
decentralized manner, each arriving node must be able to
\emph{discover} the current number of nodes in it, and the nodes to
connect to, i.e. its neighbors. We will present a technique for this
later. Before doing so, we analyze the statistical
characteristics of the resulting network topologies, and show that they
are scale-free.

As before, strong clustering is a direct consequence of the triangle
inequality in the hidden space, but we have to show that
the node degree distribution follows a power law at any time instance,
as soon as the number of nodes in the network is sufficiently large.
The analysis becomes more
complicated than in Section \ref{sec:static}, because the
hyperbolic disc radius is no longer constant, but grows with the number
of nodes in the network.

To proceed, we first compute: (1)~the
probability density function $f(r,t)$ of the radial coordinate $r$ of a
randomly selected node when the number of nodes in the network is some
value $t\gg1$, and (2)~the average degree $\bar{k}(r,t)$ of nodes located
at distance $r$ from the disc center. Having
the expressions for $f(r,t)$ and $\bar{k}(r,t)$ ready, we can then tell
how these quantities scale as a function of $r$, and use
arguments similar to those in the proof of Theorem \ref{thm:one} to
find the degree distribution.  The derivation of $f(r,t)$ and $\bar{k}(r,t)$
is in the Appendix. Below we present the final expressions and
show that the degree distribution is a power law.

Let $R(t)$ be the hyperbolic disc radius when the number of nodes in
the network is $t$. According to the model, $R(t)=\frac{1}{\alpha}
\ln\frac{t}{c}$. The approximate expression for the node
radial density $f(r,t)$ is:
\begin{equation}
\label{eq:f_r_t}
f(r,t) \approx \alpha^2 (R(t)-r) e^{\alpha(r-R(t))},
\end{equation}
leading to:
\newtheorem{lem}{\textbf{Lemma}}
\begin{lem}
\label{lem:one}
For $r<R(t)$ and $t\gg1$, $f(r, t) \sim e^{\alpha r}$.
\end{lem}

For $\alpha > \frac{1}{2}$, the average node degree as a function of $r$ and $t$ is,
approximately:
\begin{equation}
\label{eq:bar_k_r_t_final}
\bar{k}(r,t) \approx t \left\{P(r)+\frac{1-e^{-\alpha(R(t)-r)}}{\alpha (R(t)-r)} \left( G(r)-P(r) \right) \right\},
\end{equation}
where:
\begin{eqnarray}
\label{eq:G_r}
\nonumber G(r)=\left(1-\frac{2\alpha^2}{\pi(\alpha-\frac{1}{2})^2}+(1-\frac{2\alpha}{\pi(\alpha-\frac{1}{2})})\alpha r \right)e^{-\alpha r}\\
+\frac{2\alpha^2}{\pi(\alpha-\frac{1}{2})^2}e^{-\frac{1}{2}r},
\end{eqnarray}
\begin{equation}
\label{eq:P_r}
P(r) = \frac{2\alpha}{\pi(\alpha-\frac{1}{2})}e^{-\frac{1}{2}r}-\frac{1}{2(\alpha-\frac{1}{2})}e^{-2\alpha\ln\frac{\pi}{2}}e^{-\alpha r},
\end{equation}
and the limit $\alpha \to \frac{1}{2}$ is well defined.~\footnote{For $\alpha <\frac{1}{2}$ Equation (\ref{eq:bar_k_r_t_final}) does not hold. See \cite{tech_report} for more details.}
The equations above lead to:
\begin{lem}
\label{lem:two}
For $\alpha  \geq \frac{1}{2}$ and $t\gg1$, $\bar{k}(r,t) \sim e^{-\frac{1}{2} r}$.
\end{lem}

Taken together the two lemmas above result in:
\begin{thm}
\label{thm:two}
The described growing model with $\alpha \geq \frac{1}{2}$ produces graphs with a power
law node degree distribution $P(k, t) \sim k^{-\gamma}$, where $\gamma=2\alpha+1$.
\begin{proof}
The proof is similar to that of Theorem \ref{thm:one}. For $t$ sufficiently large, by Lemma \ref{lem:two},  $\bar{k}(r, t) \sim e^{-\frac{1}{2} r}$. Hence, $\bar{r}(k, t) \sim -2 \ln k$.  By Lemma \ref{lem:one}, $f(r,t) \sim e^{\alpha r}$, approximately. Therefore, $P(k, t) \approx f(\bar{r}(k,t), t) |\bar{r}'(k,t)| \sim k^{-\gamma}$, with $\gamma=2\alpha+1$.
\end{proof}
\end{thm}

The average node degree $\bar{k}(t)=\int_{0}^{R(t)}\bar{k}(r,t)f(r,t)dr$ is:
\begin{equation}
\label{eq:k_t}
\small\bar{k}(t) \approx 2\alpha c \left(e^{(\alpha-\frac{1}{2})R(t)} C_1-\alpha R(t)^2 C_2-R(t)C_3-C_4\right),
\end{equation}
where $C_1=\frac{2\alpha^3}{\pi(\alpha-\frac{1}{2})^3
  (2\alpha-\frac{1}{2})}$,
$C_2=\frac{2\alpha-\pi(\alpha-\frac{1}{2})}{2\pi(\alpha-\frac{1}{2})}$,
$C_3=\frac{\alpha}{\pi(\alpha-\frac{1}{2})^2}$,
$C_4=\frac{2}{\pi(\alpha-\frac{1}{2})}\left(\frac{\alpha}{2(\alpha-\frac{1}{2})^2}+1\right)$,
and the limit $\alpha \to \frac{1}{2}$ is again well defined. Using
Equation (\ref{eq:k_t}) we can thus choose constant $c$
to set the average node degree to a target value
in synthetic networks grown to a target size $t$. For
fixed $c$ and $\alpha \to \frac{1}{2}$ the first exponential term in
Equation (\ref{eq:k_t}) vanishes, and $\bar{k}(t)$ grows
very slowly with $t$, as a function of $\ln t$, since
$R(t)=\frac{1}{\alpha} \ln\frac{t}{c}$. This property is desirable, as in
practical applications, we want the average node degree to depend
weakly on the system size. Interestingly, by Theorem
\ref{thm:two}, $\alpha = \frac{1}{2}$ yields degree exponent
$\gamma=2$, which as we have seen in Section \ref{sec:static},
also maximizes the efficiency of greedy forwarding.

\begin{figure}[!ht]
\begin{center}
\begin{tabular}{cc}
\begin{minipage}{1.6in}
\includegraphics [width=1.6in, height=1.0in]{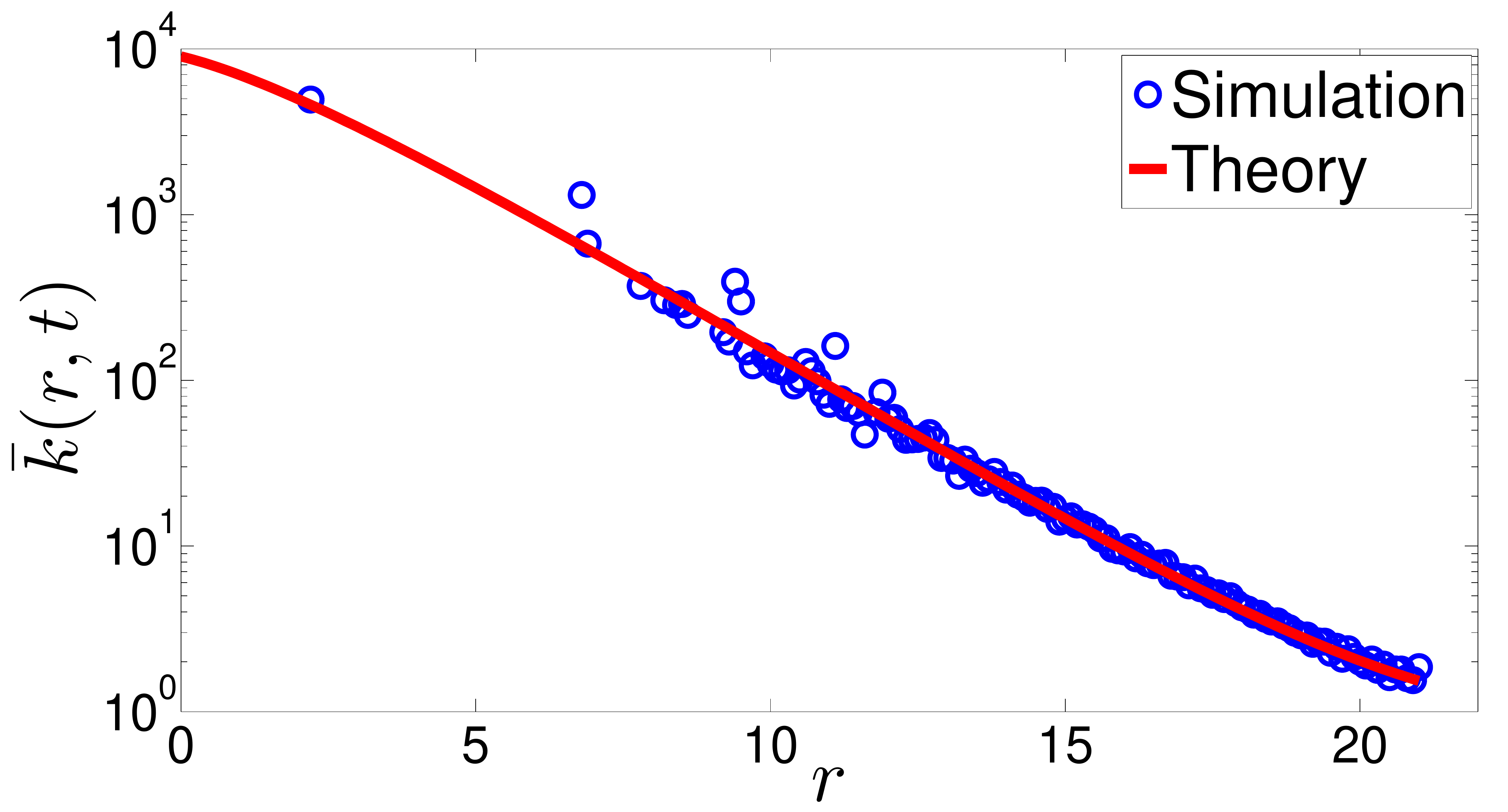}
\end{minipage} &
\begin{minipage}{1.6in}
\includegraphics [width=1.6in, height=1.02in]{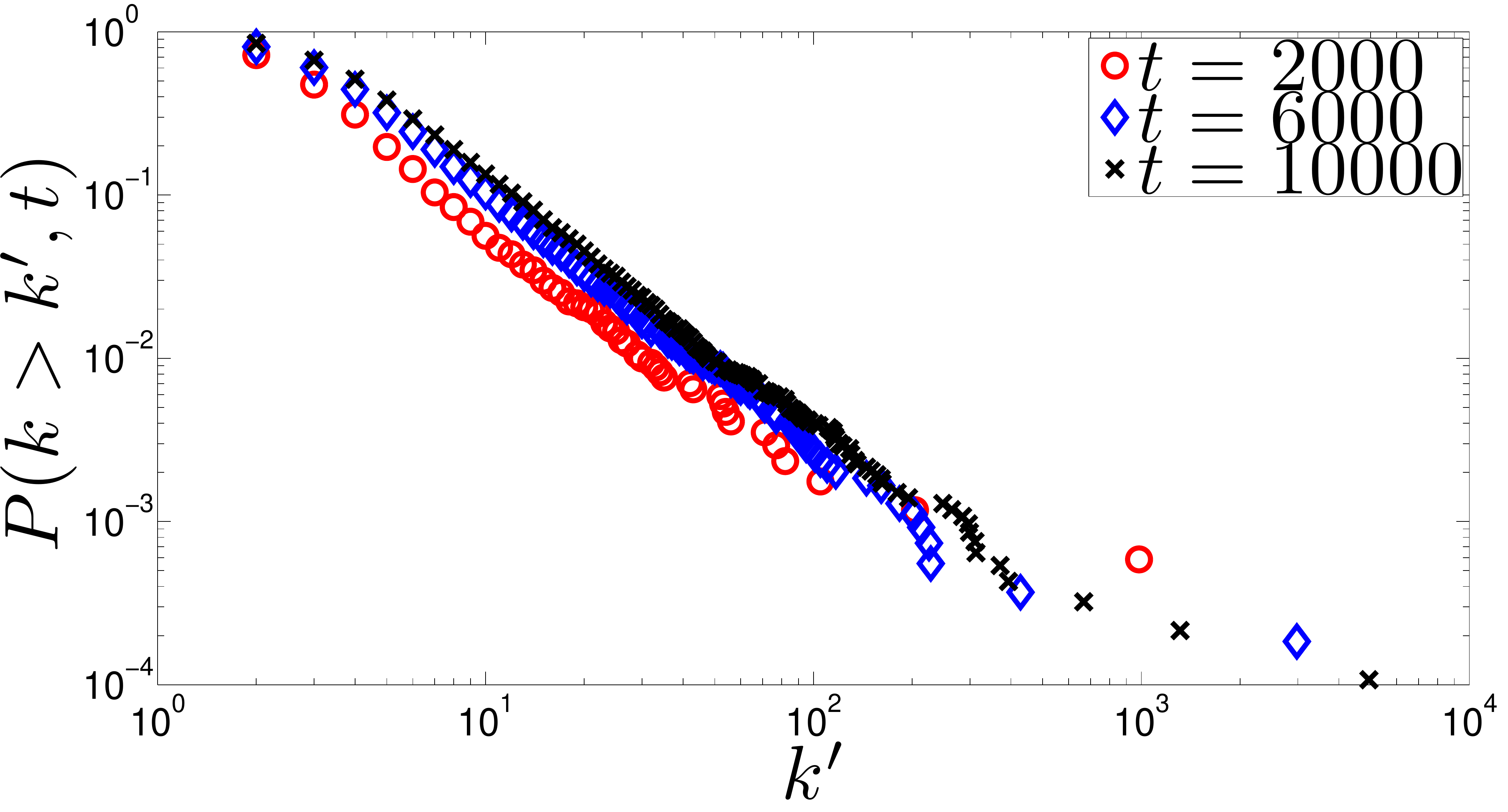}
\end{minipage} \\
(a) & (b)
\end{tabular}
\caption{(a) $\bar{k}(r, t)$ at $t=10000$, and (b) degree ccdf at various $t$.}
\label{fig:growing_verify}
\end{center}
\end{figure}

In Figure \ref{fig:growing_verify}, we check the accuracy of our
analytic predictions. The figure is for a synthetic network growing
with parameters $\alpha=0.75$, i.e., target $\gamma=2.5$, and $c=0.0014$,
i.e., target $\bar{k}(t)=6.5$ when $t=10000$.
Figure \ref{fig:growing_verify}(a) shows, in semi-log scale, that
Equation (\ref{eq:bar_k_r_t_final}) closely matches
simulations. Figure \ref{fig:growing_verify}(b) shows, in log-log
scale, the node degree ccdf as the network grows, i.e., at
various $t$.  The ccdf approximately follows a power law with the slope,
i.e., exponent $\gamma-1$,~\footnote{If pdf $P(k)$ is a power law with
  exponent $\gamma$, then ccdf $P(k>k')$ is a power law with exponent
  $\gamma-1$.} virtually independent of $t$, as Theorem
\ref{thm:two} predicts.  The exponent $\gamma$ is approximately $2.5$.  We
further observe that the ccdf is slightly shifting to the right,
indicating that the average node degree in the network grows
slowly. At $t=10000$ the average degree is $6.5768$, which is very
close to our target value. Similar results hold for other parameter
values.

\subsection{Decentralized implementation}

As mentioned earlier, to build a network in a decentralized manner,
each arriving node must be able to discover its neighbors and the current number of
nodes in the network. Below we describe an
efficient and simple greedy algorithm implementing these tasks.
This algorithm is just an example, and other techniques and
optimizations are possible.

In a nutshell, the algorithm operates as follows.
Each arriving node first contacts a random node
currently in the network, which acts as a \emph{bootstrap node}.
The bootstrap node
then sends an \emph{exploration packet} to the network. Each hop that
receives the packet writes in it its id and coordinates, as well as
the ids and coordinates off all its neighbors, and then forwards
the packet to its \emph{highest
  degree neighbor} that has not seen the packet before. The process
terminates when all neighbors of a node have seen the packet. Below
we describe the process in more detail.

The exploration packet starts from the bootstrap node and keeps a list
of the node ids it has visited, denoted by $L_{V}$. It also keeps a
list of node ids along with their corresponding coordinates, denoted
by $L_{C}$.  Each node that receives the packet records its own id
into $L_V$, and its id and coordinates into $L_C$. Further, it also
records the id and the corresponding coordinates of each of its
neighbors into $L_C$.  The node then selects from its neighbors that
are not included in $L_V$, the one with the maximum degree, and
forwards the packet to it. The process terminates when all neighbors
of a node are listed in $L_V$, in which case the exploration packet is
sent back to the bootstrap node.  The list $L_C$ is then given to the
arriving node.

This process, called \emph{search utilizing high degree nodes}, is
very efficient in power law graphs.  In particular, for degree
exponents $2<\gamma <3$, the exploration packet can discover a
large percentage of nodes in the graph along with their coordinates
(recorded in $L_C$), by traversing only a small number of hops $\sim
N^{2-\frac{4}{\gamma}}$ (recorded in $L_V$), see Chapter $13$ in
\cite{BoSchu02-book}. Having an estimate of the number of nodes, the
arriving node can compute the current hyperbolic disc radius, and in
turn, its own coordinates. Knowing the hyperbolic disc
radius, the coordinates of the nodes, and their ids, it can also
compute to which nodes it should connect.

One possible modification of the above basic technique is to impose
an upper bound on the size of the $L_C$ list. Once such a bound is
reached, the current $L_C$ list is
returned to the bootstrap node, and then cleared in the exploration
packet. This extension adds control on the maximum size that the
exploration packet can have.

\subsection{Greedy forwarding}

We now evaluate via simulation the performance of greedy forwarding in
highly dynamic conditions, with random node arrivals and departures.
Our setup is as follows.

Without loss of generality, time is slotted. During each time slot, a
new node arrives w.p.\ $p=0.1$, and each node currently in the network
departs w.p.\ $q=10^{-5}$. Initially the network consists of $0$ nodes.
An arriving node joins the network according to our growing model. It
also discovers the current number of nodes in it and their coordinates
according to the procedure described earlier.  In our experiments
below the exploration packet discovers $95\%$ of nodes in
the network on average, by traversing only $1.5\%$ of all nodes.
If all
neighbors of a node depart, the node re-initiates the join process to
reconnect to the network.  To ensure that the network remains
connected, we assume that the first $t_{start}=200$ nodes
never depart. According to our growing model, these nodes will have
high degrees. High degree nodes are required to
maintain connectivity in scale-free graphs \cite{BoSchu02-book}.
See also the discussion in Section \ref{sec:applications}.

The average number of nodes in the network grows and stabilizes at the
steady state value
$\bar{t}_{steady}=\frac{p}{q}+t_{start}=10200$.~\footnote{
  In the steady state we still have node arrivals and departures, but the
  network does not grow on average. Indeed, if $\bar{t}(s)$ is the
  average number of nodes in some time slot $s$, excluding the first
  $t_{start}$ nodes, and $\bar{t}(s+1)$ is the corresponding number in
  slot $s+1$, then $\bar{t}(s+1)=\bar{t}(s)(1-q)+p$, leading to the steady state
  value of $\bar{t}=\lim_{s \to \infty} \bar{t}(s)=\frac{p}{q}$. Adding
  $t_{start}$ to this last quantity yields $\bar{t}_{steady}$.}
Note that $\frac{t_{start}}{\bar{t}_{steady}} \approx 2\%$.
We set the system parameters to $\alpha=0.5$ and
$c=0.01$, which according to our earlier analysis,
correspond to the average
node degree of $\bar{k}({\bar{t}_{steady})}=6.5$ and degree exponent
$\gamma=2$. Recall from Section \ref{sec:static} that $\gamma=2$
maximizes the efficiency of greedy forwarding, and makes
the average degree depend logarithmically on the network
size.

\begin{figure*}
    \centerline{
        \subfigure[Average node degree.]{\includegraphics[width=1.8in, height=1.15in]{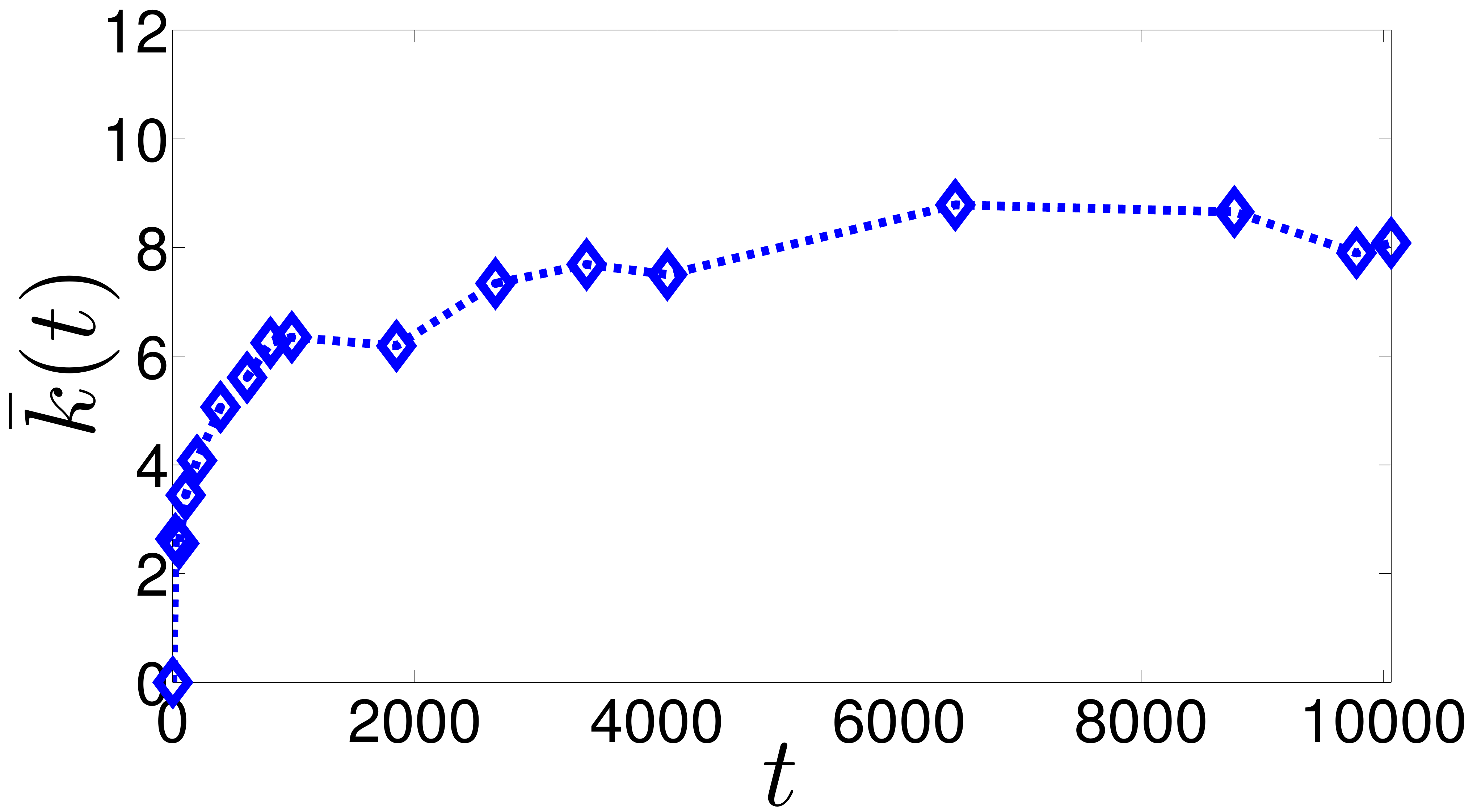}}\hfill
        \subfigure[Degree ccdf.]{\includegraphics[width=1.8in, height=1.2in]{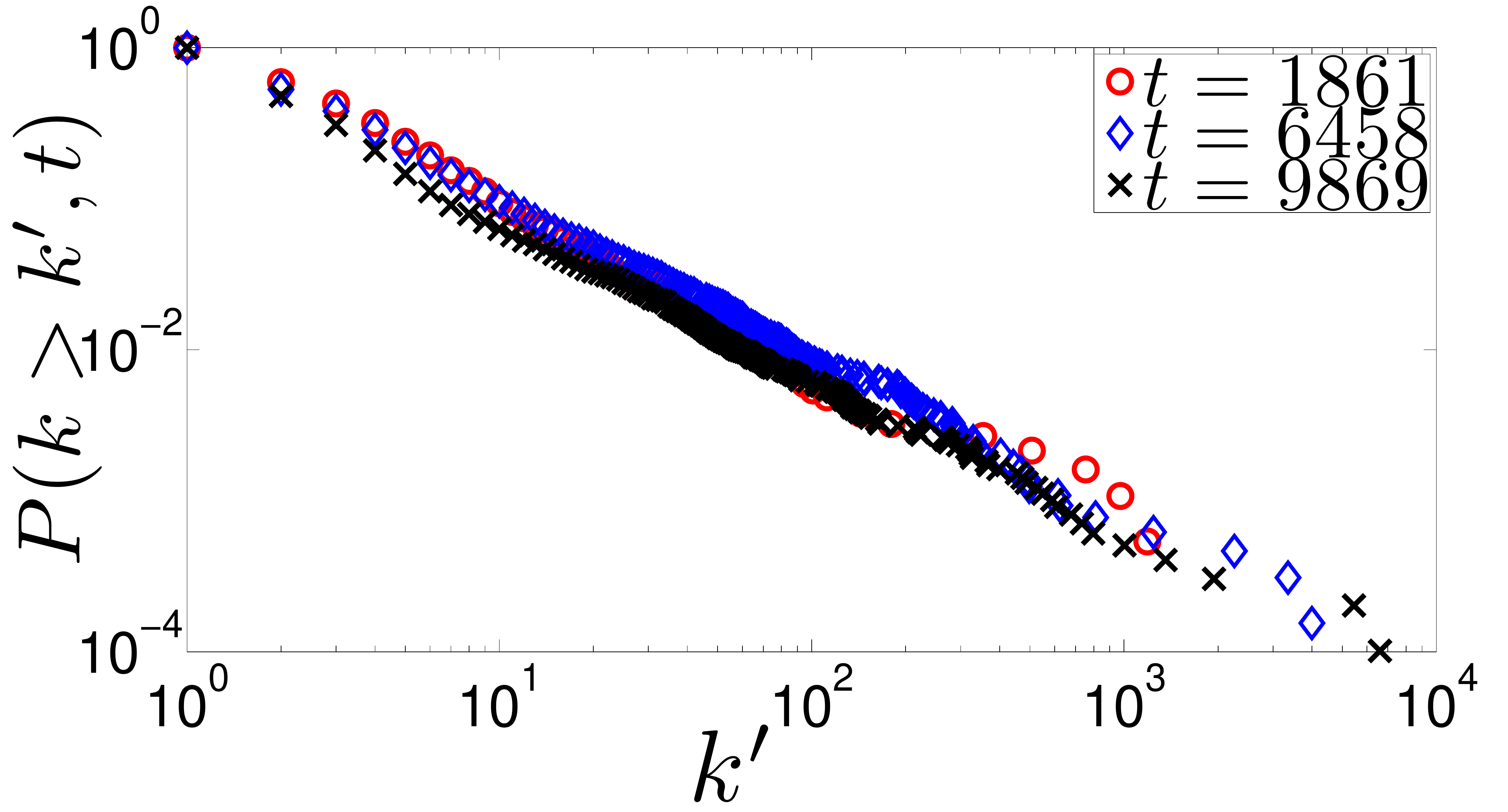}}\hfill
        \subfigure[Success ratio.]{\includegraphics[width=1.8in, height=1.2in]{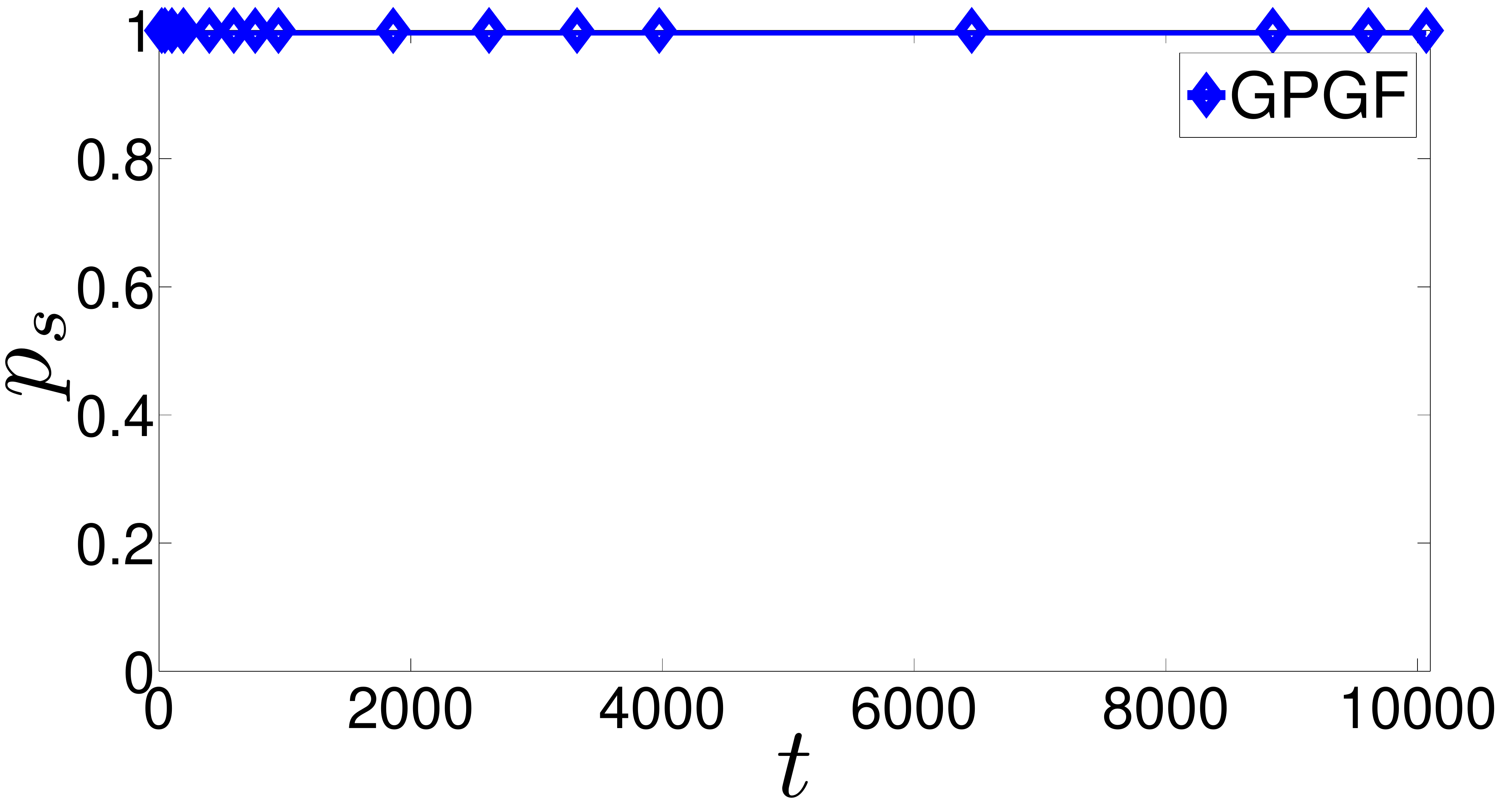}}\hfill
        \subfigure[Stretch.]{\includegraphics[width=1.8in, height=1.2in]{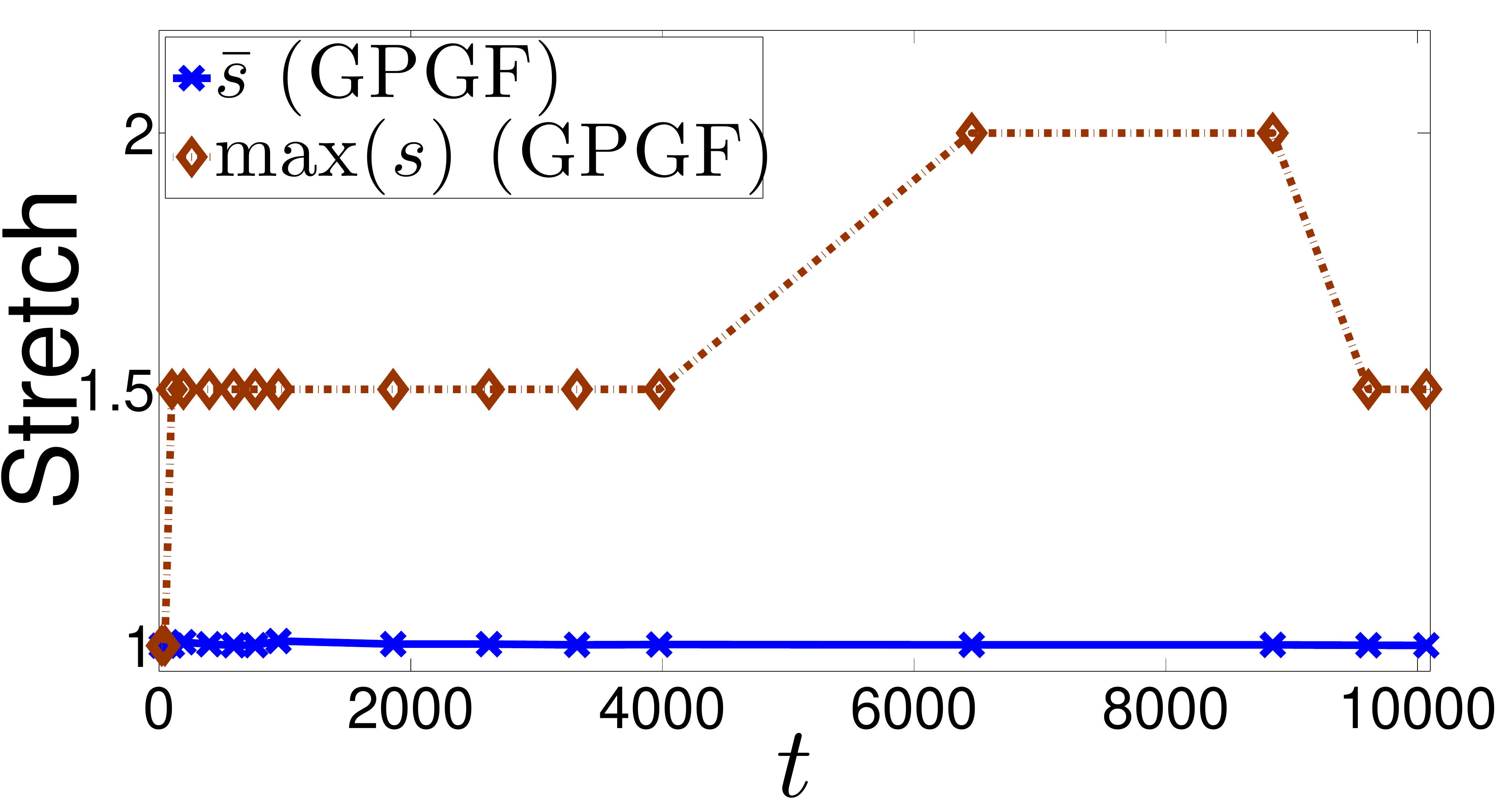}}
    }
    \caption{Dynamic networks.
    \label{fig:gpgf_performance}
    }
\end{figure*}

Figure \ref{fig:gpgf_performance}(a) shows the average node degree
$\bar{k}(t)$ as a function of the current number of nodes $t$ in the
network, until and when we reach steady state.  The average degree
grows initially and then stabilizes above, but close to, our target value of
$6.5$.~\footnote{The observed discrepancy is due to that the
network is not growing exactly according to the model. Instead,
it grows \emph{on average}. The exact analysis accounting for the
specifics of the node arrival and departure stochastic processes
is beyond the scope of this paper.}
Figure \ref{fig:gpgf_performance}(b) shows the node degree ccdf.
The degree exponent is $\gamma \approx 2$, and it does not change
as the network grows, as expected.

As mentioned in Section \ref{sec:static}, we expect
simple GF modifications to achieve better performance
than our OGF and MGF strategies, even in highly dynamic
network conditions. We check these expectations with
the \emph{Gravity-Pressure Greedy Forwarding}
algorithm (GPGF) from \cite{crovella09infocom}, described below.

\underline{Gravity-Presure Greedy Forwarding (GPGF)}. Each packet
carries a bit to indicate whether the packet is in \emph{Gravity} or
\emph{Pressure} forwarding mode.  The packet starts in Gravity mode,
where the forwarding procedure is exactly the same as in our OGF
algorithm. However, if the packet reaches a local minimum, it is not
dropped as in OGF. Instead, it first records the distance of the local
minimum to the destination, which we call \emph{current local-minimum
  distance}, and then enters Pressure mode. In Pressure mode, the
packet maintains a list of the nodes in the network it has visited
since it entered this mode, and the number of visits to each node. A
node that receives the packet determines all neighbors that
the packet has visited the least number of times, selects among
those the one with the minimum distance to the destination, and forwards
the packet to this neighbor. This process continues until the packet either reaches
the destination or a node whose distance to the destination is smaller
than the current local-minimum distance. In the latter case, the
packet switches back to and continues in the Gravity mode.

Regardless of underlying space geometry, the success ratio of
GPGF is guaranteed to be always $p_s=1$ \cite{crovella09infocom},
which we confirm in
Figure \ref{fig:gpgf_performance}(c).  What is \emph{not} guaranteed
by the algorithm is the \emph{stretch}, which can be enormous, as in
the worst case a packet can visit \emph{all} nodes in the
network to find its destination. However, in Figure
\ref{fig:gpgf_performance}(d) we see that GPGF's stretch is
exceptionally low in our networks. In particular, we see
that the average stretch $\bar{s}$ remains extremely close to $1$,
while the maximum stretch max$(s)$ never exceeds $2$. As before, this
remarkable efficiency is due to the congruency between scale-free
network topology and underlying hyperbolic geometry, which persists
even in highly dynamic conditions.

%% file: discussion.tex
\section{Discussion}
\label{sec:applications}

In this section we discuss potential practical applications of our results.
One of such applications is overlay network construction \cite{p2psurvey, overlay_routing}.
Indeed, the idea of using an
underlying geometry to guide the forwarding process is similar in
spirit to Distributed Hash Table (DHT) overlay architectures
\cite{p2psurvey}. In overlay networks, messages searching for
data content are usually greedily forwarded, based on some distance
metric, to the node in the network responsible for the data,
usually the node that is closer to the data in terms of the distance
metric. The main objective is to have a low average node degree, and
a short average hop-length of search paths \cite{p2psurvey}.
Both metrics should grow slowly with the
number of nodes in the network $N$. The best
efficiency with $O(\ln\ln N)$-long search paths, and $O(1)$ average node degree,
can be achieved only in scale-free networks since only in scale-free
networks do shortest paths grow as $\sim \ln \ln N$, independently
of the average node degree \cite{BoSchu02-book}.
Therefore, given that greedy paths are approximately shortest in our
networks, a property we proved in \cite{ultrashort},
our results can be used for overlay network construction
and routing \cite{overlay_routing} to improve routing/search efficiency.

However, by no means do we propose a solution which is better
than all existing overlay architectures in \emph{all} aspects.
For example, it is often desirable
that all nodes in a network have similar degrees,
which is not the case with scale-free networks.
Another concern may be that while scale-free
networks are robust to random node failures/departures, they are
vulnerable to failures of the highest degree nodes, resulting
in network disconnection \cite{BoSchu02-book}. To address this concern,
one may consider, for example, hybrid architectures with
resilient super-nodes (the high degree nodes) \cite{hybrid}, or use
techniques similar to those in \cite{phenix} to ensure recovery when
such nodes fail. However, as we have demonstrated in the previous
section, it is enough to ensure that high degree nodes corresponding
to only $2\%$ of the total steady state number of nodes, do not fail.

Our results
also suggest that forwarding information through existing scale-free networks, such as the Internet, should be possible
without global topology knowledge and associated routing overhead, which is one of the most serious
scaling limitations with the existing Internet architecture \cite{iab-raws-report-phys}.
In other words, our results lay the groundwork for potentially moving to an Internet where
forwarding can take place without routing.

%% file: conclusion.tex
\section{Conclusion}
\label{sec:conclusion}

We have shown that scale-free network topologies naturally emerge
from, and are congruent with, hyperbolic geometries.  This congruency
can be used to efficiently guide the forwarding process to find
destinations with $100\%$ success probability, following almost
optimal shortest paths, even in highly dynamic networks.
Our findings complement the important work in
\cite{kleinberg00-nature}, and can have several practical applications
in overlay network construction for improving routing/search
performance.

There are several interesting directions for future work. One is to
explore other techniques for decentralized implementation.  Another is
the design of greedy forwarding strategies that could also be used for
improving other network performance metrics, e.g., strategies that
could avoid congestion areas, perform load-balancing, and so on.

Finally, our results suggest that forwarding information through existing scale-free
networks, e.g., the Internet, should be possible without routing overhead.
Therefore, one of the  most interesting, yet challenging, future work
directions is the following \emph{inverse problem}: Can we embed any
\emph{real} scale-free network, e.g., the Internet topology, into
a hyperbolic space, so that we can greedily forward through this embedding
with similar efficiency?  How can each node compute its
coordinates in the space \emph{having no global knowledge of the network
topology}, so that the resulting embedding is congruent with
the space, requiring no coordinate and routing updates
even if the network is highly dynamic?

%% file: paper.bbl

%% file: appendix.tex
\renewcommand{\appendixname}{Appendix -- Growing Model Analysis}
\appendix
Here we consider a growing network at some time instance where the number of nodes in it is some value $t$,
and derive the expressions for the node radial density $f(r,t)$ and the average node degree $\bar{k}(r,t)$.
We number each node by its order of arrival. Recall that there are no node departures, and that each
node $i \geq 1$ that arrives to the system computes a hyperbolic disc radius
$R(i)=\frac{1}{\alpha} \ln\frac{i}{c}$. The radial coordinate $r \in [0, R(i)]$ of node $i$ is distributed according to
the density  $f(r|R(i)) \approx \alpha e^{\alpha(r-R(i))}$.

We start with $f(r,t)$. Let $\{R(1)...R(t)\}$ be the sequence of the hyperbolic disc radii that nodes $\{1...t\}$ compute on their arrival, and let $R$ be the random variable representing the computed disc radius of a randomly selected node from $\{1...t\}$. It is easy to see that for $i \in \{1...t\}$, $P(R \leq R(i))=\frac{i}{t}=\frac{c e^{\alpha R(i)}}{ce^{\alpha R(t)}}=e^{\alpha(R(i)-R(t))}$.

We treat $R$ as a continuous random variable. As we see in Section \ref{sec:dynamic} this does not affect the accuracy of the predictions.
Therefore, if we denote by $F(R,t)=e^{\alpha(R-R(t))}$ the distribution function of the computed disc radius $R$, then the probability density function of $R$, denoted by $f(R,t)$,  is $f(R,t)=\alpha e^{\alpha (R-R(t))}$
and it is obtained by differentiating $F(R,t)$ w.r.t. $R$.  Since, given the value of $R$ a node computes its radial coordinate $r$ according to $f(r|R) \approx \alpha e^{\alpha(r-R)}$ and $r \leq R \leq R(t)$, we can write $f(r,t)=\int_{r}^{R(t)} f (r|R) f (R,t) dR$.  Performing the integration, we get Equation (\ref{eq:f_r_t}).

We now proceed with $\bar{k}(r,t)$. Its computation is rather long, and we omit it for brevity, see \cite{tech_report}.
We can break $\bar{k}(r,t)$ into two parts. The first part $\bar{k}_{init}(r,t)$ is the initial average degree of nodes with radial coordinate $r$, i.e., the average
number of nodes already in the network, to which the node connects upon its arrival. The second part $\bar{k}_{new}(r,t)$ is the average number of new
connections to nodes at $r$, coming from new nodes arriving to the system after. Clearly, $\bar{k}(r,t)=\bar{k}_{init}(r,t)+\bar{k}_{new}(r,t)$.

We first compute $\bar{k}_{init}(r,t)$. Suppose that the node at $r$ computed a disc radius equal to $R$ when it arrived. According to our model, the node then connected to
all other nodes in the network, for which the hyperbolic distance to it was  $d \leq R$. The average number of these nodes, denoted by $\bar{k}_{init}(r|R)$,
can be computed \cite{tech_report}, and its approximate expression for $\alpha \geq \frac{1}{2}$  is $\bar{k}_{init}(r|R) \approx ce^{\alpha R} G(r)$, where $G(r)$ as given
 by Equation (\ref{eq:G_r}).

To obtain $\bar{k}_{init}(r,t)$, we now have to remove the condition on $R$ from $\bar{k}_{init}(r|R)$ by accounting for all possible $R$ with $r \leq R \leq R(t)$.
To do so, we need the conditional  probability
 density for $R$, given that a node's radial coordinate is $r$, denoted by $f(R,t|r)$.  Given  $f(r|R)$, $f(R,t)$ and $f(r,t)$,
 we have $f(R,t|r)=\frac{f(r|R) f(R,t)}{f(r,t)}$, and $\bar{k}_{init}(r,t) = \int_{r}^{R(t)} \bar{k}_{init}(r|R) f (R,t|r) dR$.

We now proceed with $\bar{k}_{new}(r,t)$. Let $N_{new}$ be the number of new nodes that arrived to the system after a node with radial coordinate $r$.
Clearly, the computed hyperbolic disc radius $R'$ of a new such node satisfies $r \leq R' \leq R(t)$. According to our model, the new node connects to the node at $r$ only if the hyperbolic distance
 to it is $d \leq R' $.
  In \cite{tech_report} we show that if $\alpha \geq \frac{1}{2}$ the probability that a new node connects to the node at $r$ is approximately independent of the exact value of $R'$ and depends only on $r$.
 This probability, denoted by $P(r)$, can be easily computed, and is given by Equation (\ref{eq:P_r}).

 Now, given that the node at $r$ was the $i^{th}$ node arrival, $N_{new}=t-i$.  If $\bar{k}_{new}(r,t|i)$  denotes the average number of new connections to node $i$, we can write $\bar{k}_{new}(r,t | i) \approx (t-i)P(r)=(t-ce^{\alpha R(i)})P(r)$. To find $\bar{k}_{new}(r,t)$, we need to remove the condition on the index $i$. In other words, we need to account again for all $R$ with $r \leq R \leq R(t)$. The conditional density of $R$, $f(R,t|r)$, was computed earlier. Therefore, $\bar{k}_{new}(r,t) \approx \int_{r}^{R(t)}(t-ce^{\alpha R})P(r)  f (R,t|r)  dR$.
 Adding $\bar{k}_{init}(r,t)$ and $\bar{k}_{new}(r,t)$, and performing the integration, we get $\bar{k}(r,t)$ given by Equation (\ref{eq:bar_k_r_t_final}).

%% file: paper.bbl
\begin{thebibliography}{10}
\providecommand{\url}[1]{#1}
\csname url@samestyle\endcsname
\providecommand{\newblock}{\relax}
\providecommand{\bibinfo}[2]{#2}
\providecommand{\BIBentrySTDinterwordspacing}{\spaceskip=0pt\relax}
\providecommand{\BIBentryALTinterwordstretchfactor}{4}
\providecommand{\BIBentryALTinterwordspacing}{\spaceskip=\fontdimen2\font plus
\BIBentryALTinterwordstretchfactor\fontdimen3\font minus
  \fontdimen4\font\relax}
\providecommand{\BIBforeignlanguage}[2]{{%
\expandafter\ifx\csname l@#1\endcsname\relax
\typeout{** WARNING: IEEEtran.bst: No hyphenation pattern has been}%
\typeout{** loaded for the language `#1'. Using the pattern for}%
\typeout{** the default language instead.}%
\else
\language=\csname l@#1\endcsname
\fi
#2}}
\providecommand{\BIBdecl}{\relax}
\BIBdecl

\bibitem{iab-raws-report-phys}
D.~Meyer, L.~Zhang, and K.~Fall, Eds., \emph{RFC4984}.\hskip 1em plus 0.5em
  minus 0.4em\relax The Internet Architecture Board, 2007.

\bibitem{NaGro05}
V.~Naumov and T.~Gross, ``Scalability of routing methods in ad hoc networks,''
  \emph{Performance Evaluation}, vol.~62, pp. 193--209, 2005.

\bibitem{p2psurvey}
E.~K. Lua, J.~Crowcroft, M.~Pias, R.~Sharma, and S.~Lim, ``A survey and
  comparison of peer-to-peer overlay network schemes,'' \emph{IEEE
  Communications survey and tutorial}, vol.~7, no.~2, 2005.

\bibitem{TraMi69}
J.~Travers and S.~Milgram, ``An experimental study of the small world
  problem,'' \emph{Sociometry}, vol.~32, pp. 425--443, 1969.

\bibitem{kleinberg00-nature}
J.~Kleinberg, ``Navigation in a small world,'' \emph{Nature}, vol. 406, p. 845,
  2000.

\bibitem{newman03c-review}
M.~E.~J. Newman, ``The structure and function of complex networks,'' \emph{SIAM
  Rev}, vol.~45, no.~2, pp. 167--256, 2003.

\bibitem{DorMen-book03}
S.~N. Dorogovtsev and J.~F.~F. Mendes, \emph{Evolution of Networks: From
  Biological Nets to the {Internet} and {WWW}}.\hskip 1em plus 0.5em minus
  0.4em\relax Oxford: Oxford University Press, 2003.

\bibitem{kleinberg07infocom}
R.~Kleinberg, ``Geographic routing using hyperbolic space,'' in \emph{INFOCOM},
  2007.

\bibitem{crovella09infocom}
A.~Cvetkovski and M.~Crovella, ``Hyperbolic embedding and routing for dynamic
  graphs,'' in \emph{INFOCOM}, 2009.

\bibitem{overlay_routing}
A.~Nakao, L.~Peterson, and A.~Bavier, ``Scalable routing overlay networks,''
  \emph{SIGOPS Oper. Syst. Rev.}, vol.~40, no.~1, 2006.

\bibitem{kleinberg06-review}
J.~Kleinberg, ``Complex networks and decentralized search algorithms,'' in
  \emph{ICM}, 2006.

\bibitem{ShaTa04b}
Y.~Shavitt and T.~Tankel, ``On the curvature of the {Internet} and its usage
  for overlay construction and distance estimation,'' in \emph{INFOCOM}, 2004.

\bibitem{KraLe06}
R.~Krauthgamer and J.~R. Lee, ``Algorithms on negatively curved spaces,'' in
  \emph{FOCS}, 2006.

\bibitem{AbBa07}
I.~Abraham, M.~Balakrishnan, F.~Kuhn, D.~Malkhi, V.~Ramasubramanian, and
  K.~Talwar, ``Reconstructing approximate tree metrics,'' in \emph{PODC}, 2007.

\bibitem{JoLoBo08}
E.~Jonckheere, P.~Lohsoonthorn, and F.~Bonahon, ``Scaled {Gromov} hyperbolic
  graphs,'' \emph{J. Graph Theory}, vol.~57, no.~2, 2008.

\bibitem{BoSchu02-book}
S.~Bornholdt and {H. G. Schuster (Edts.)}, \emph{Handbook of Graph and
  Networks: From the Genome to the {Internet}}.\hskip 1em plus 0.5em minus
  0.4em\relax Berlin: Wiley-VCH, 2002.

\bibitem{BarAlb99}
A.-L. Barab\'{a}si and R.~Albert, ``Emergence of scaling in random networks,''
  \emph{Science}, vol. 286, pp. 509--512, 1999.

\bibitem{BridsonHaefliger99-book}
M.~R. Bridson and A.~Haefliger, \emph{Metric Spaces of Non-Positive
  Curvature}.\hskip 1em plus 0.5em minus 0.4em\relax Berlin: Springer-Verlag,
  1999.

\bibitem{GiNe02}
M.~Girvan and M.~E.~J. Newman, ``Community structure in social and biological
  networks,'' \emph{Proc Natl Acad Sci USA}, vol.~99, pp. 7821--7826, 2002.

\bibitem{WatDoNew02}
D.~J. Watts, P.~S. Dodds, and M.~E.~J. Newman, ``Identity and search in social
  networks,'' \emph{Science}, vol. 296, pp. 1302--1305, 2002.

\bibitem{ClMo08}
A.~Clauset, C.~Moore, and M.~E.~J. Newman, ``Hierarchical structure and the
  prediction of missing links in networks,'' \emph{Nature}, vol. 453, pp.
  98--101, 2008.

\bibitem{Gromov07-book}
M.~Gromov, \emph{Metric Structures for Riemannian and Non-Riemannian
  Spaces}.\hskip 1em plus 0.5em minus 0.4em\relax Boston: Birkh{\"a}user, 2007.

\bibitem{tech_report}
F.~Papadopoulos, D.~Krioukov, M.~Boguna, and A.~Vahdat, ``Greedy forwarding in
  dynamic scale-free networks embedded in hyperbolic metric spaces,''
  \emph{CAIDA Technical Report number tr-2009-02}, 2009,
  \url{http://www.caida.org/publications/papers/2009/tr-2009-02/}.

\bibitem{WatStr98}
D.~J. Watts and S.~H. Strogatz, ``Collective dynamics of ``small-world''
  networks,'' \emph{Nature}, vol. 393, pp. 440--442, 1998.

\bibitem{routeviews}
``{University of Oregon RouteViews Project},''
  \url{http://www.routeviews.org/}.

\bibitem{dimes}
``The {DIMES} project,'' \url{http://www.netdimes.org/}.

\bibitem{curvtemp}
D.~Krioukov, F.~Papadopoulos, A.~Vahdat, and M.~Boguna, ``Curvature and
  temperature of complex networks,'' \emph{Phys. Rev. E}, vol.~80, no.~3, 2009.

\bibitem{ultrashort}
M.~Bogu{\~{n}}\'{a} and D.~Krioukov, ``Navigating ultrasmall worlds in
  ultrashort time,'' \emph{Phys Rev Lett}, vol. 102, p. 058701, 2009.

\bibitem{hybrid}
C.~Gkantsidis and M.~Mihail, ``Hybrid search schemes for unstructured
  peer-to-peer networks,'' in \emph{INFOCOM}, 2005.

\bibitem{phenix}
R.~H. Wouhaybi and A.~Campbell, ``Phenix: supporting resilient low-diameter
  peer-to-peer topologies,'' in \emph{INFOCOM}, 2004.

\end{thebibliography}
